\documentclass[preprint2]{aastex}

\usepackage{epsfig}
\usepackage{textcomp}

\shortauthors{Hatziminaoglou et al.}
\shorttitle{SDSS Quasars in the SWIRE EN1 Field: Properties and SEDs}

\begin{document}

\newcommand\mums{\textmu m }
\newcommand\mum{\textmu m}

\title{Sloan Digital Sky Survey Quasars in the SWIRE ELAIS N1 Field: Properties and Spectral Energy Distributions}

\author{E. Hatziminaoglou\altaffilmark{1}, I. P\'{e}rez-Fournon\altaffilmark{1},
M. Polletta\altaffilmark{2}, A. Afonso-Luis\altaffilmark{1}, A. Hern\'{a}n-Caballero\altaffilmark{1},
F.M. Montenegro-Montes\altaffilmark{1}, C. Lonsdale\altaffilmark{2,3}, C.K. Xu\altaffilmark{3}, 
A. Franceschini\altaffilmark{4}, M. Rowan-Robinson\altaffilmark{5}, T. Babbedge\altaffilmark{5},
H.E. Smith\altaffilmark{2}, J. Surace\altaffilmark{3}, D. Shupe\altaffilmark{3}, 
F. Fang\altaffilmark{3}, D. Farrah\altaffilmark{3}, S. Oliver\altaffilmark{6},
E.A. Gonz\'{a}lez-Solares\altaffilmark{7}, S. Serjeant\altaffilmark{8}}

\altaffiltext{1}{Instituto de Astrofisica de Canarias, C/ V\'{\i}a L\'{a}ctea s/n, E-38200 La Laguna, 
Tenerife, Spain; evanthia@iac.es}
\altaffiltext{2}{Center for Astrophysics and Space Sciences, University of California, San Diego, La Jolla, 
CA 92093-0424, USA}
\altaffiltext{3}{Infrared Processing and Analysis Center, California Institute of Technology, Pasadena, CA 91125, USA}
\altaffiltext{4}{Dipartimento di Astronomia, Universita di Padova, Vicolo Osservatorio 5, 35122 Padua, Italy}
\altaffiltext{5}{Astrophysics Group, Blackett Laboratory, Imperial College London, London SW7 2BW, UK}
\altaffiltext{6}{Astronomy Centre, Department of Physics and Astronomy, University of Sussex, Falmer,
Brighton BN1 9QJ, UK}
\altaffiltext{7}{Institute of Astronomy, University of Cambridge, Madingley Road, Cambridge CB3 0HA, UK}
\altaffiltext{8}{Centre for Astrophysics and Planetary Science, School of Physical Sciences, 
University of Kent, Canterbury, Kent CT2 7NR, UK}


\begin{abstract}
We present a mid-infrared analysis of 35 quasars with spectroscopic redshifts selected from the Spitzer Wide-area 
InfraRed Extragalactic Survey (SWIRE). We discuss their optical and mid-infrared (MIR) colors, and show that these 
quasars occupy well defined regions in MIR color-color space. We examine the issue of type-I AGN candidate
selection in detail and propose new selection methods based on mid-IR colors.
The available multi-band data allows us to construct two new, well-sampled 
quasar templates, covering wavelengths from the ultraviolet to the MIR.

\end{abstract}


\section{Introduction}

Efforts to understand the physics that drive the activity in active galactic nuclei (AGN) have been underway for 
decades, and have recently seen rapid progress thanks to improvements in both instrument design and theoretical modeling. 
Observationally, new facilities have provided important pieces of information, which when combined with advances in 
modeling have allowed a reasonably clear picture of the emission mechanisms in AGN to emerge. The optical continuum 
(up to 1 \mum) of type-I AGN, is seen to be a power law, which can be explained by the existence of an accretion disk 
around a central supermassive black hole. At longer wavelengths, one starts observing emission thought to arise from 
a dusty torus, which, depending on the geometry, peaks at wavelengths between 15 and 100 \mum. This emission is thought 
to be the reprocessed emission of the UV/optical radiation from the accretion disk by the particles composing the torus, 
namely silicate and graphite grains (e.g. \citealt{granato94}; \citealt{efstathiou95}; \citealt{nenkova02}).

Studies with the Infrared Space Observatory (ISO) showed that the mid-infrared (MIR) window can be used to distinguish 
AGN from starbursts \citep{rigo99,laurent00,farrah02,farrah03}. Furthermore, recent work conducted on a sample consisting 
of 25 quasars detected by the European Large Area ISO Survey (ELAIS) at 15 \mums \citep{afonso04} showed the potential 
of combined optical and MIR observations in the construction of quasar samples, and in achieving a better understanding 
of the physical properties of AGN. These studies also showed, however, the need for deeper multi-band observations in 
the MIR in order to refine physical constraints on the torus models and unified schemes. 

With the advent of Spitzer it is now possible to obtain MIR photometry for very large samples of galaxies. In particular, 
the Spitzer Wide-area InfraRed Extragalactic Survey (SWIRE; \citealt{lonsdale03}; \citealt{lonsdale04}) offers an unprecedented 
opportunity to study the MIR properties of AGN thanks to both the large areal coverage ($\sim 50$ deg$^2$) and to the number 
of bands in which the fields are being observed (IRAC bands 3.6, 4.5, 5.8 and 8 \mums and MIPS bands 24, 70 and 160 \mum).
In this paper we investigate properties of a complete optically selected spectroscopically confirmed type-I quasar sample 
from the Sloan Digital Sky Survey (SDSS), consisting of 35 objects. The paper is structured as follows. Section \ref{sample} 
describes the MIR, optical and near-IR data available for this sample; these properties are then used in Section \ref{midselection} 
in order to determine type-I AGN candidate selection criteria from multi-color data. Section \ref{seds} discusses the SEDs 
of the objects. Section \ref{bhmass} deals with the Black Hole (BH) mass and IR luminosity estimates. Notes on individual 
objects are given in Section \ref{notes}. Finally, Section \ref{discuss} presents a discussion of the results.

\section{The Quasar Sample in the various Wavelengths}
\label{sample} 
The MIR data used here are taken from the SWIRE ELAIS N1 field, and were obtained in February 2004 with both IRAC and MIPS. For the 
purposes of this work only the four IRAC bands (referred to as IRAC1, IRAC2 etc throughout) and MIPS 24 \mums (hereafter MIPS24) are used. The 
SWIRE catalogs we use throughout this work were processed by the SWIRE collaboration. Details about the data can be found in 
\cite{lonsdale04}, \cite{surace04} and \cite{shupe04}.

The Sloan Digital Sky Survey has validated and made publicly available its Data Release 2 (DR2; \citealt{abazajian04}), partially 
covering the SWIRE EN1 field. DR2 comprises 32241 quasars with redshift $<$2.3 and 3791 quasars with redshift $>$2.3. A total of 35 
spectroscopically confirmed quasars lie within the 3.5 deg$^2$ of SWIRE EN1 covered by the SDSS DR2 spectroscopic release.
Two of the objects only have 24 \mums detections as they fall in areas at the very border of the field and are not covered by IRAC,
while four others have detections in all four IRAC bands but do not have MIPS 24 \mums observations due to the failure of some MIPS 
scans during the MIPS campaign. None of the non-detections are drop-outs, instead all are due to the inhomogeneous coverage of the 
field at its edges. Their $i$-magnitudes reach 19.1 for objects with redshifts typically less than 2.3 and go up to a magnitude 
deeper for higher redshifts \citep{richards02}. Table \ref{tabquasars} shows the (optical) positions, redshifts, SDSS AB magnitudes
and MIR fluxes and errors (in mJy) for the 35 quasars composing our sample. The sequence number in the first column
in used hereafter as an identifier for the objects.

The quasar sample was matched with the 2MASS all sky catalog and near-IR counterparts were found for six objects, including the highest
redshift object of the sample at $z=3.653$. Six of the objects also have ELAIS 15 \mums counterparts (\citealt{vaccari04}; \citealt{rowan04}; \citealt{afonso04}). 
Details can be found in Section \ref{notes}.

\begin{figure*}[h!]
\centerline{
\psfig{figure=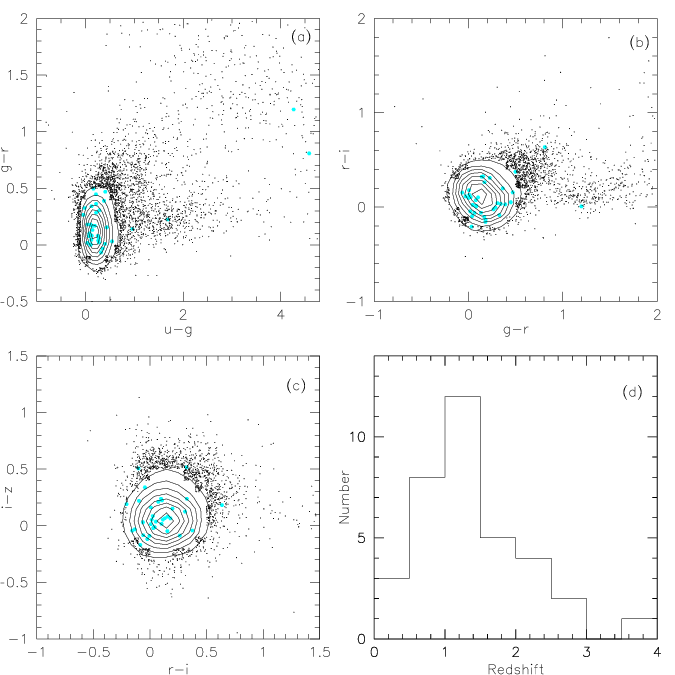}}
\caption{Optical (AB) colors (panels a, b and c) and redshift distribution (panel d) 
of the quasar sample (cyan filled circles). Black contours and black points indicate the location of the
$\sim$ 16700 spectroscopically confirmed quasars of SDSS DR1.}
\label{figoptprop}
\end{figure*}

Fig. \ref{figoptprop} (a), (b) and (c) show the optical colors (in AB) of the 35 quasars (cyan filled circles) superposed on the 16710 quasars of the SDSS Data Release 1 (DR1; \citealt{abazajian03}), occupying the same regions 
of color space. Fig. \ref{figoptprop} (d) shows the redshift distribution of our sample, which peaks at slightly lower redshift as 
compared to the redshift distribution of the DR1 quasar catalog \citep{schneider03}.

\begin{figure*}[h!]
\centerline{
\psfig{figure=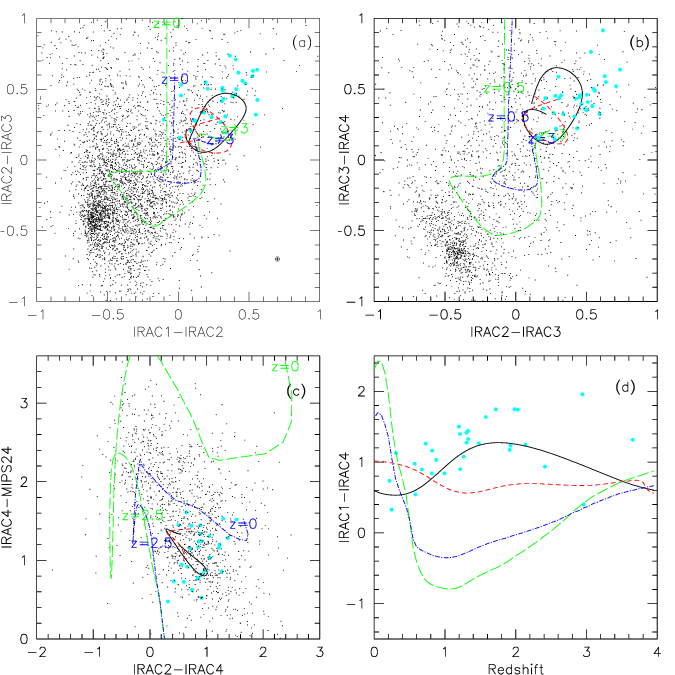}}
\caption{MIR colors of the quasar sample (filled cyan circles). The black points
are the objects in the bandmerged catalog of $tile$ 3\_1 (see text for details).
The black solid, red short-dashed, green long-dashed and blue dashed-dotted 
evolutionary tracks correspond 
to a type-I quasar, a red quasar, Arp220 and Sc templates, respectively
(from \citealt{polletta04}). For more than 90\% of all objects the error bars are smaller than that 
shown on the right bottom corner of panel (a).}
\label{figirprop}
\end{figure*}

The sample's MIR colors, defined as -2.5 $\times$ log(flux$_i$/flux$_{i+1}$)
with $i$ and $i+1$ two consecutive bands, are
shown in Fig. \ref{figirprop} (a), (b) and (c), superposed on the bandmerged catalog (see \citealt{surace04}) 
extracted from a small region (10\%, plotted for clarity) of the SWIRE EN1 field. The solid, short-dashed, long-dashed and dashed-dotted
lines correspond to evolutionary tracks of a type-I quasar, a red quasar, Arp220 and Sc templates respectively (\citealt{polletta04}). The quasars' 
colors are almost independent of their redshift due to the essentially power-law nature of their SEDs; objects of very different redshift can 
find themselves very close together in this diagram. The Arp220 and Sc templates were chosen because they are those that pass the closest to the 
quasar locus. Quasars are redder than the main bulk of the population, at least up to $\sim$ 8 \mum , but then start mixing completely with the 
rest of the objects. Both type-I and red AGN populate the redder part of the color diagrams, their very red colors being due to torus emission.
Fig. \ref{figirprop}(d) shows the (IRAC1 -- IRAC4) color as a function of redshift and compares it to the evolutionary tracks for the various 
types of objects (linestyle coding same as in the other panels of Fig. \ref{figirprop}). These well defined regions in the MIR color-color space 
can be used as a standalone criterion for type-I AGN candidate selection when optical information is unavailable, and as additional constraints 
to the optical selection and photometric redshift determination.

\begin{figure*}[h!]
\centerline{
\psfig{figure=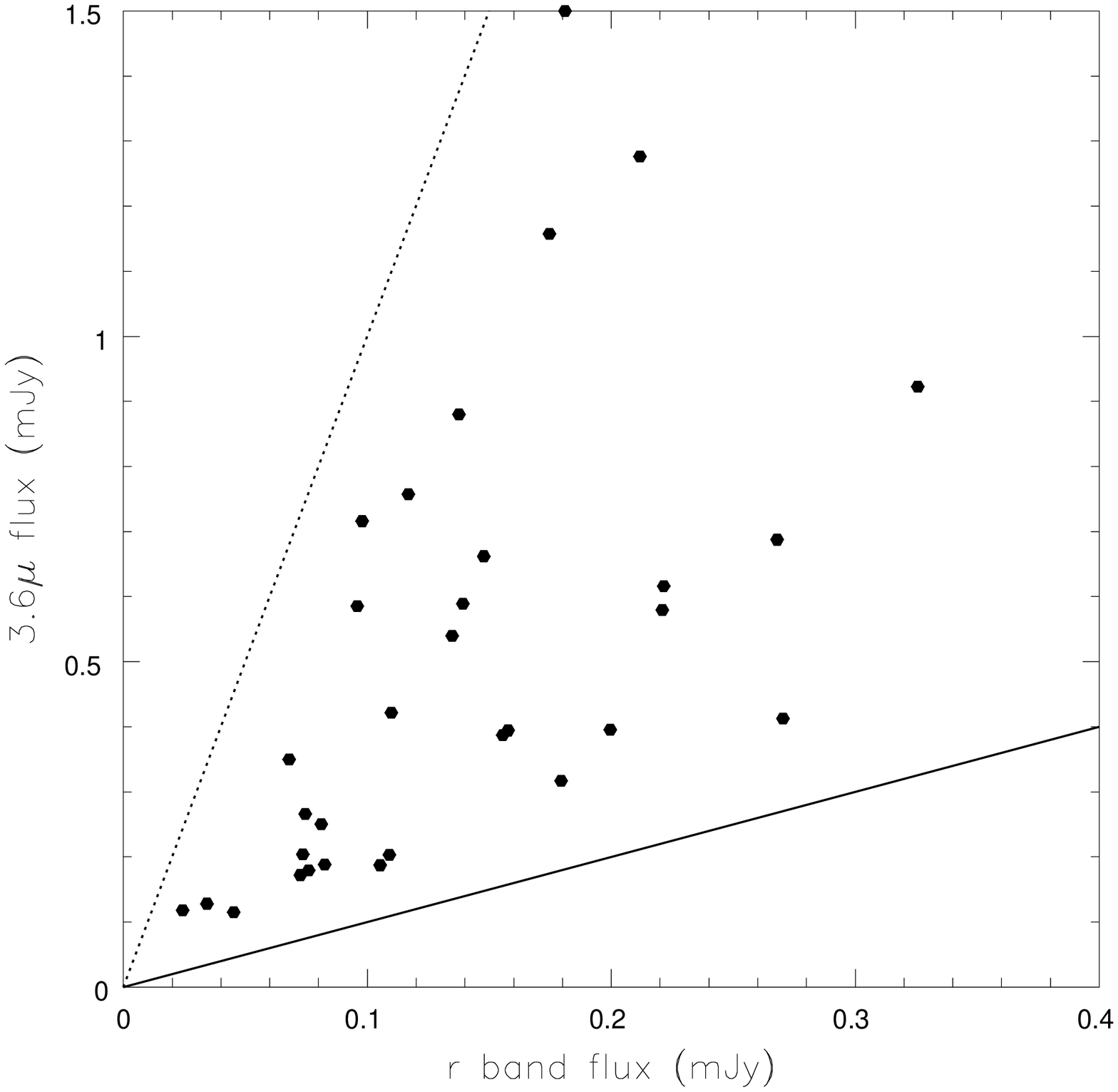,width=8cm}
\psfig{figure=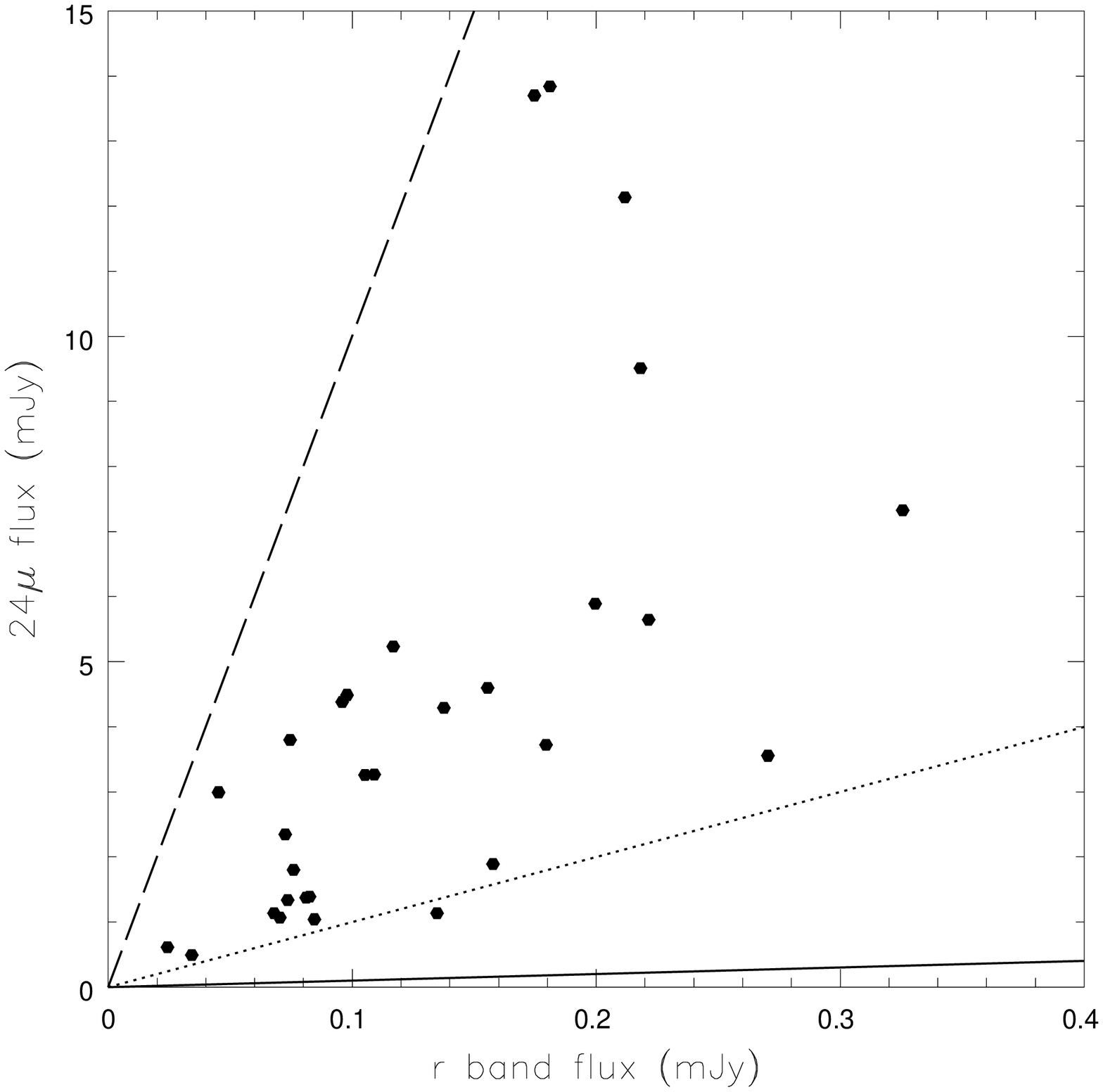,width=8cm}}
\caption{3.6 \mums (left panel) and MIPS 24 \mums (right panel) versus optical 
($r$-band) flux shown for the objects with IRAC1 and MIPS 24 \mums 
counterparts, respectively. The solid, dotted and dashed lines denote constant
IR-to-optical flux ratios, with values 1, 10 and 100, respectively.
The errors on the IR fluxes are smaller than the plotting symbols and therefore
cannot be seen on the figures.}

\label{figoptir}
\end{figure*}

Models for the IR and submm SEDs of normal, star-burst and active galaxies show that quasars have, on average, MIR fluxes some 10 to 100 times larger 
than their optical fluxes \citep{rowan01}. Fig. \ref{figoptir} compares the $r$-band flux to the IRAC1 flux at 3.6 \mums (left panel) and 
MIPS24 flux at 24 \mums (right panel) for the objects with IRAC and MIPS counterparts. The fluxes confirm the model predictions, showing that optically
bright quasars tend to have brighter (M)IR counterparts. The solid, dotted and dashed lines denote constant IR-to-optical flux ratios, with values 
1, 10 and 100, respectively.

\section{Quasar Candidate Selection in the MIR}
\label{midselection}
In this section we discuss type-I AGN candidate selection methods. No candidate list is included here
as the aim is to propose some selection criteria, describe their advantages and drawbacks and the contribution
from MIR photometry, and the color properties of the candidate sample as a whole rather than for individual objects. 
A similar discussion but proposing different selection criteria can be found in \cite{lacy04}, who present 
AGN candidates in the Spitzer First Look Survey. For an independent analysis of the location of AGN in MIR 
color space see also \cite{haas04}.

The MIR band-merged catalog in SWIRE EN1 was associated with the five-band optical catalog derived from the 
Isaac Newton Telescope Wide Field Survey (WFS; \citealt{mcmahon01}). The SWIRE EN1 and INT WFS fields do not have 
a 100\% overlap; WFS covers only $\sim $ 7 deg$^2$ of the SWIRE EN1 field, down to a limiting AB $r$-band magnitude 
of 24.1. The final bandmerged catalog consists of $\sim$293000 objects, with at least one WFS and one Spitzer
detection each. Details about the optical data and the bandmerged catalogs can be found in \cite{surace04}.

In optical surveys the use of a morphological pre-selection make main sequence stars to be the most common 
contaminants of the quasar candidates samples. With the addition of MIR data, star/quasar separation is straightforward, 
as stars occupy well defined regions in the IRAC color space due to the Rayleigh-Jeans regime in their SEDs 
(see \citealt{surace04} for a detailed discussion), separate from the regions populated by quasars. In MIR selected 
quasar samples the major contaminants will instead be normal and starburst galaxies, of both low ($z\lesssim 0.6$)
and high ($\sim$3) redshift (Fig. \ref{figirprop}). However, MIR selection can also reveal type-II AGN that would 
mostly be missed if only optical selection is used. 

\begin{figure}[h!]
\centerline{
\psfig{figure=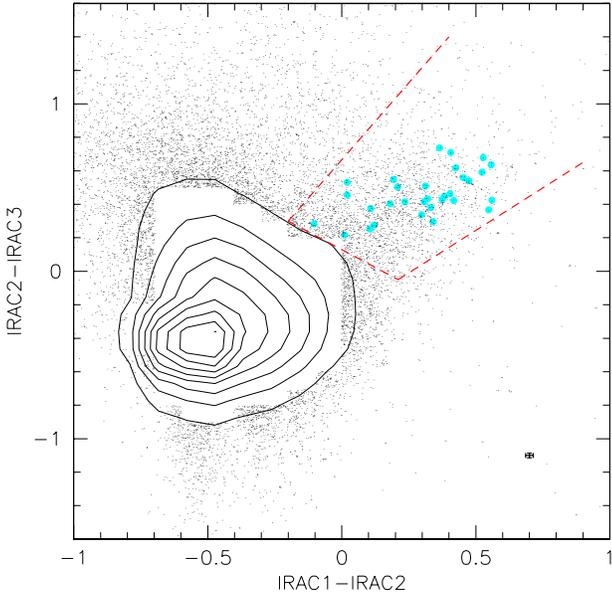,width=8.5cm}}
\caption{(IRAC1 -- IRAC2) versus (IRAC2 -- IRAC3) diagram showing the location of
the 31 out of 35 quasars with IRAC1, IRAC2 and IRAC3 counterparts occupy (cyan filled circles).
The black contours and points show objects present in the $\sim$7 deg$^2$ where WFS overlaps with
the SWIRE EN1 field. The red dashed lines delineate the region of quasar candidates.
For more than 90\% of all objects the error bars are smaller than that
shown on the right bottom corner of the figure.}
\label{figqsosel}
\end{figure}

The well defined region our quasar sample occupies can be used to constrain the color space in which
quasar candidates are selected. Fig. \ref{figqsosel} shows the location of the 35 SWIRE-SDSS quasars
in a (IRAC1 -- IRAC2) versus (IRAC2 -- IRAC3) diagram, superposed with all the sources with an $r$-band counterpart 
present in the $\sim$7 deg$^2$ where WFS overlaps with the SWIRE EN1 field. The red dashed lines delineate the region 
of quasar candidates, with equations:
$$0.35 \times (IRAC1 - IRAC2) + 0.41 \times (IRAC2 - IRAC3) = 0.053$$
$$1.1 \times (IRAC1 - IRAC2) - 0.6 \times (IRAC2 - IRAC3) = -0.4$$
$$(IRAC1 - IRAC2) - (IRAC2 - IRAC3) = 0.26.$$
The region defined here is only indicative, and the tolerance depends on the degree of completeness and the number of 
contaminants that are acceptable.

A total of 1870 quasar candidates populate the region inside the dashed lines. Ths number drops to 1290 when detections 
in all four IRAC channels, MIPS 24 \mums and $r$-band are required. This number rises to 2520 when only IRAC channels 1-3 
detections are needed (which are the only relevent bands for the selection criterion applied here). 

A type-I AGN sample selected based on MIR colors cannot be entirely complete, and could also suffer from contamination by
spiral and starburst galaxies at specific redshifts. When optical morphology is taken into account the resulting 
sample will be cleaner but less complete, as nearby type-I AGN will probably be excluded. Note also that the more MIR 
information one adds the more accurate the selection can become (see for instance \citealt{lacy04}), but then one is 
limited by the substantially lower sensitivity of IRAC channels 3 and 4. Type-II AGN are more difficult to identify. For an 
in depth discussion on their colors see \cite{polletta04}. Using X-ray selected AGN, \cite{franceschini04} show that the 
optical to MIR SEDs of type-II AGN can match those of normal spiral, starbursting and even passively evolving elliptical 
galaxies, while only their X-ray signature indicates the presence of nuclear activity.

\subsection{Model predictions on quasar MIR counts}\label{model}
The number of quasars in this sample, and their optical and MIR fluxes can give an estimate of the number of AGN SWIRE will detect.
All 35 quasars are at least 3 times brighter than the 5$\sigma$ limit at 5.8, 8.0 and 24 \mums and at least 25 times brighter 
at 3.6 and 4.5 \mum, have $i < 19.1$ when $z \le 2.3$ and $i < 20.1$ when $z > 2.3$. Using these limits, expected numbers 
of AGN have been calculated based on a new version of the \cite{xu03} models, including a new evolutionary model for dusty galaxies.
Galaxies with obscured AGN are those with a MIR excess in the rest frame: $f_{25\mu m}/f_{60\mu m} \geq 0.2$. Their evolution 
function is assumed to be the same as that of optical quasars, which is a pure luminosity evolution function of the form \citep{boyle00}:
\begin{equation}
L^*(z) = L^*(0) 10^{1.36z-0.27z^2} \;\; (z\leq 7)
\end{equation} 
but with $L^*$ now defined at 25\mum.

\begin{figure}
\centerline{
\psfig{figure=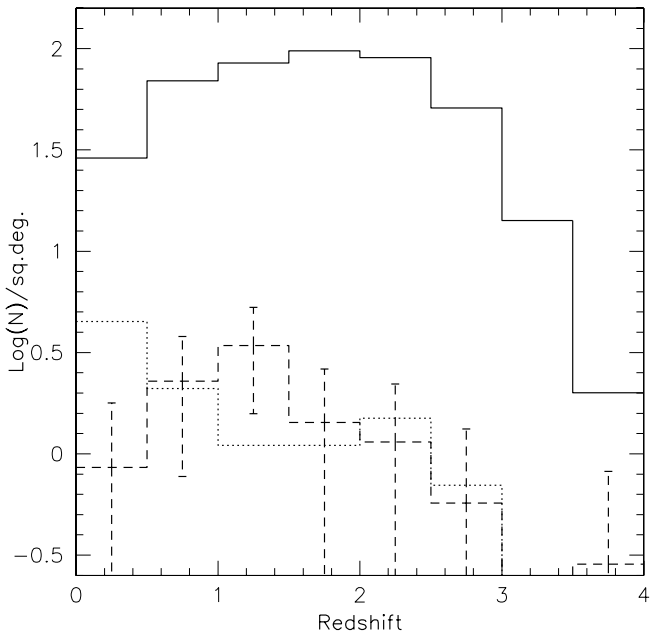,width=9cm}}
\caption{Histograms showing the number of AGN per deg$^2$
expected for the SWIRE 5$\sigma$ limits (solid line), the number of AGN expected
when the SDSS spectroscopic limits are applied (dotted line)
and the actual quasars (dashed line).}
\label{figzhist}
\end{figure}

Fig. \ref{figzhist} shows the predicted redshift distributions per deg$^2$ from this model, compared to the actual 
distribution of the 35 SDSS quasars (dashed line). The solid line corresponds to the model realisation using the SWIRE 5$\sigma$ 
limits in both IRAC and MIPS and the dotted line is the distribution predicted when the SDSS optical spectroscopic limits
are applied as well. Note that the model does not distinguish between type-I and type-II AGN.
The dashed line is the scaled redshift distribution of the 35 quasars, normalised to 1 deg$^2$.

\begin{figure*}[h!]
\centerline{
\psfig{figure=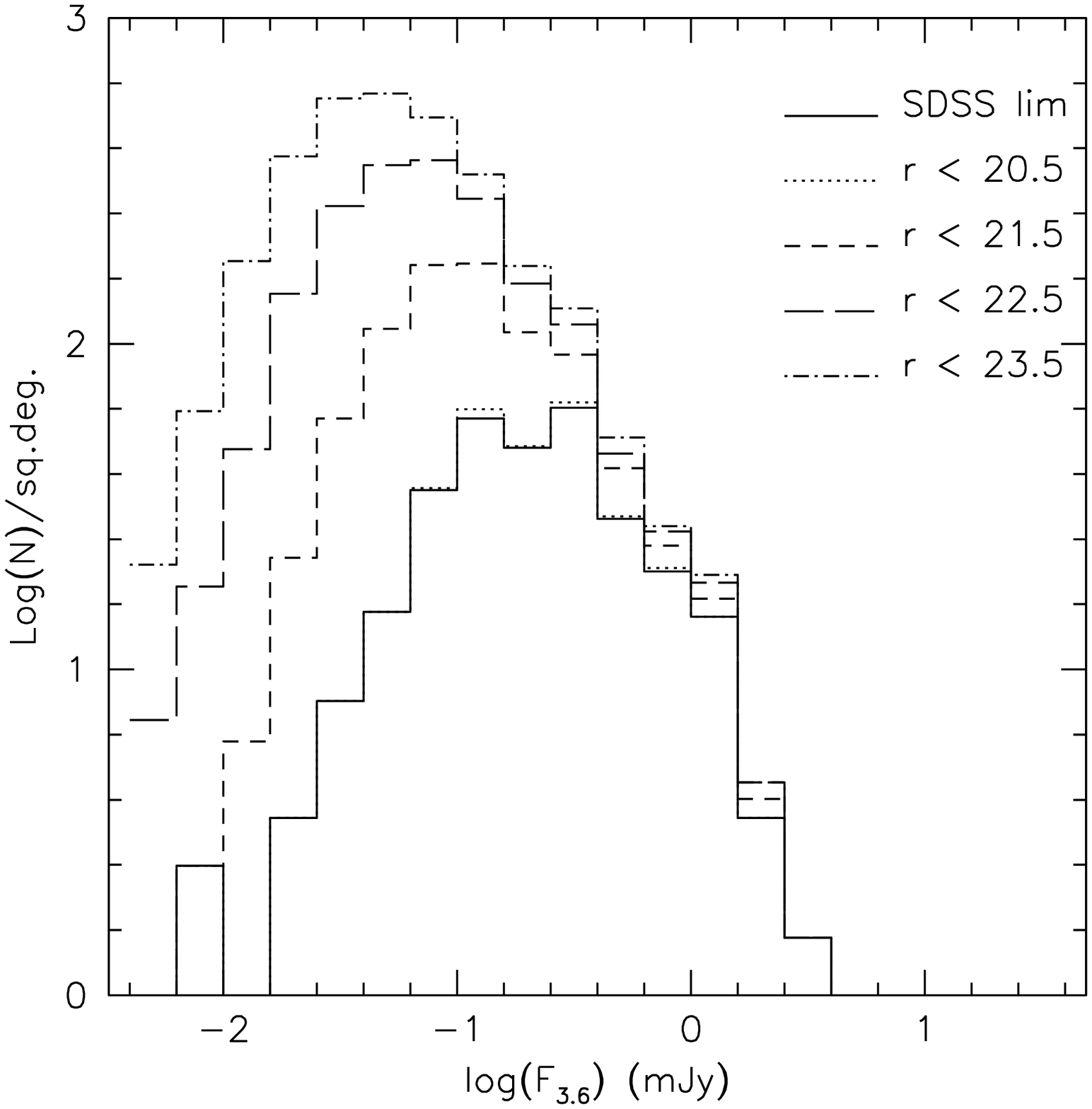,width=8cm}
\psfig{figure=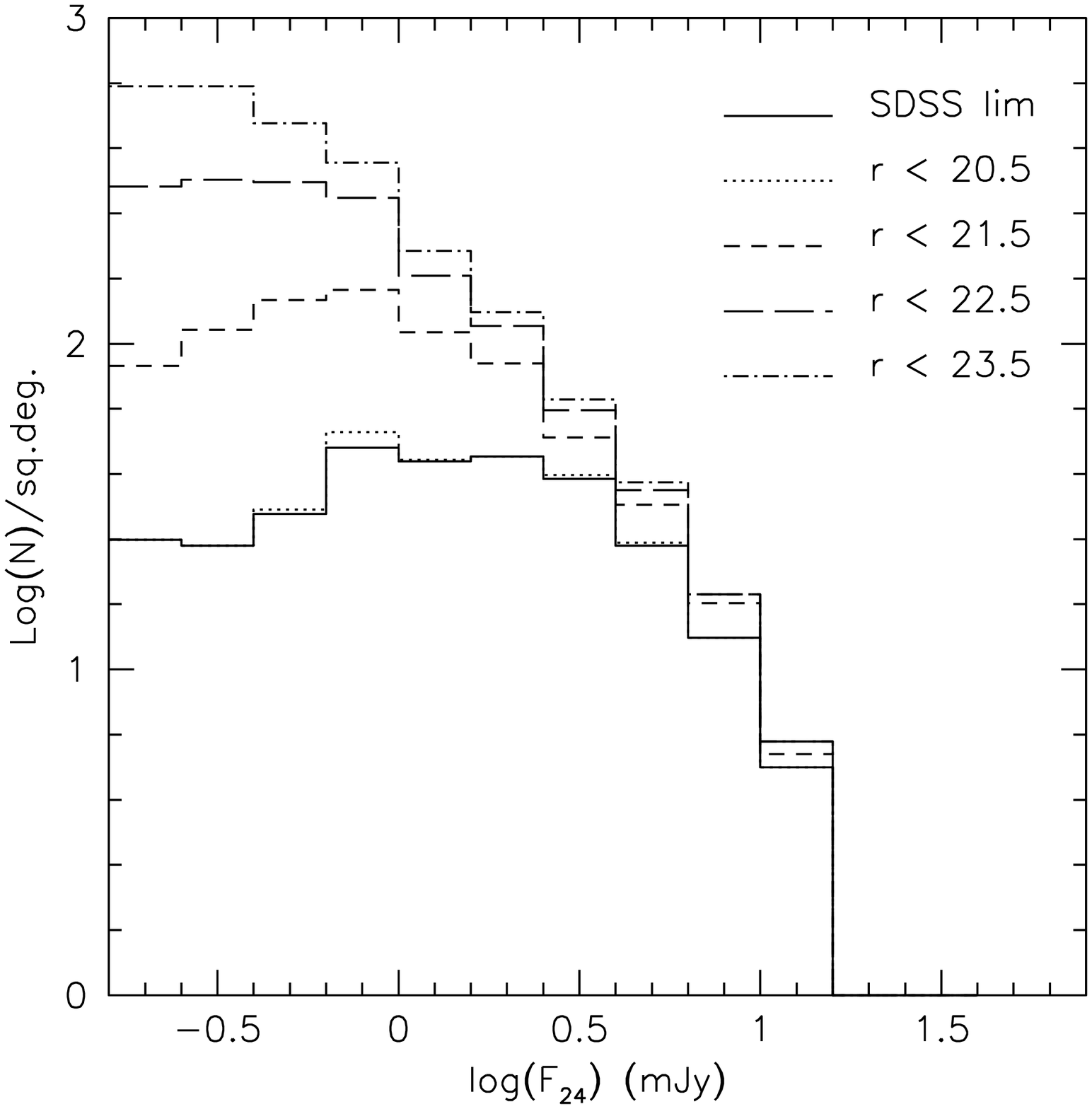,width=8cm}}
\caption{AGN number counts for various optical ($r$-band) limits predicted,
at 3.6 (left) and 24 \mums (right) fluxes. Only the respective MIR detections
are required, increasing the estimated number by a factor of almost 3.5 with respect 
to Fig. \ref{figzhist}.}
\label{figncounts}
\end{figure*}

The total number of objects predicted by the model for a 3.5 deg$^2$ area (which is the coverage of SWIRE EN1 by the SDSS DR2) 
using the SWIRE 5$\sigma$ limits and the SDSS $i$-band spectroscopic limit was 39, i.e. 11 per deg$^2$, very close to the actual 
number of quasars in our sample. The predicted redshift distribution is also close to the observed one, as can been seen by comparing 
the dotted line with the black dashed line in Fig. \ref{figzhist}. This good agreement is most probably
due to the cut in the $i$-band, which most likely excludes all type-II AGN.
The model predicts some 440 AGN (types I and II 
together) per deg$^2$ with redshifts up to 4 and detections in all IRAC bands and the MIPS 24 \mums channel down to the SWIRE MIR 
limits (when no optical constraints are imposed), with some 20 $z>3$ objects per deg$^2$. The difference between the observed and 
estimated numbers implies a large amount of incompleteness introduced in the quasar samples by optical constraints. The number of 
objects (1870 in the $\sim$7 square degrees of SWIRE EN1 covered by the WFS data) previously selected as quasar candidates is 1.6 
times smaller than the one predicted by the model. This number drops to 1.2 (i.e. 360 candidates per deg$^2$) when no optical 
counterpart is required. The model estimates for the AGN differential number counts 
(N/deg$^2$/dex) at 3.6 and 24 \mums for various optical 
magnitude limits are shown in Fig. \ref{figncounts}. Note that on these plots only the respective MIR detections are required, 
increasing the estimated number by a factor of almost 3.5 with respect to the previous model realisation where all IRAC band and 
MIPS 24 \mums are required. With very few exceptions, the quasars in our sample have much brighter 3.6 and 24 \mums fluxes
than the fluxes where the model counts peak.

The missing population consists of type-I quasars omitted from the optical selection, type-I AGN without all IRAC and MIPS detections
as well as type-II AGN that are mostly optically obscured or optically extended. In fact, \cite{lacy04} suggest that up to 50\% of 
their MIR selected AGN could be sufficiently obscured not to be observed by SDSS.

\section{Spectral Energy Distributions}\label{seds}
\begin{figure*}
\centerline{
\psfig{figure=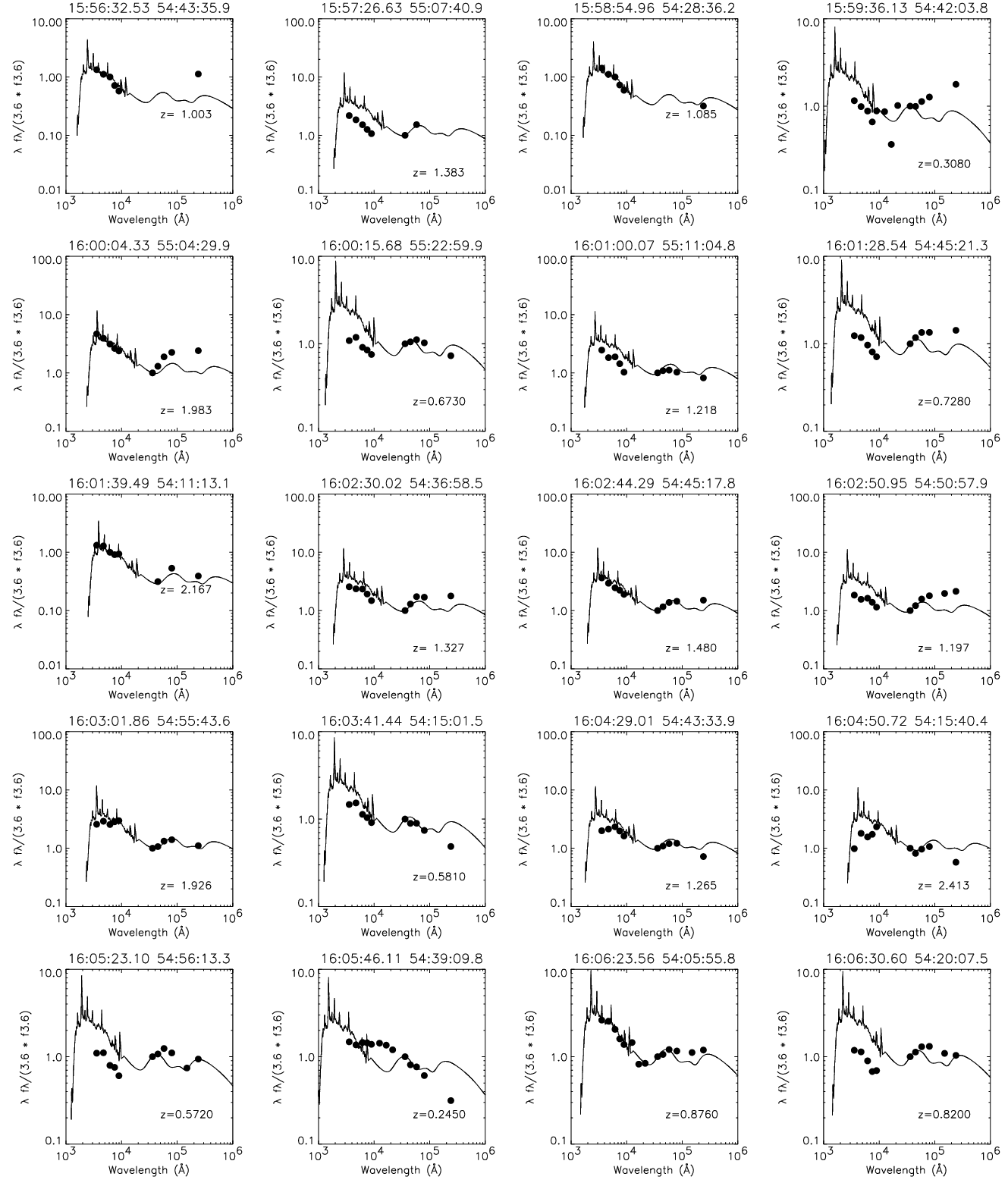}}
\caption{Observed SEDs for the 35 SWIRE-SDSS quasars (filled circles) overplotted to 
an average quasar template from a subsample of the PG survey 
described in the text, shifted to the redshift of the object.
The objects appear in the same order as in Table \ref{tabquasars}, i.e. ordered by RA.}
\label{figseds1}
\end{figure*}

\begin{figure*}
\centerline{
\psfig{figure=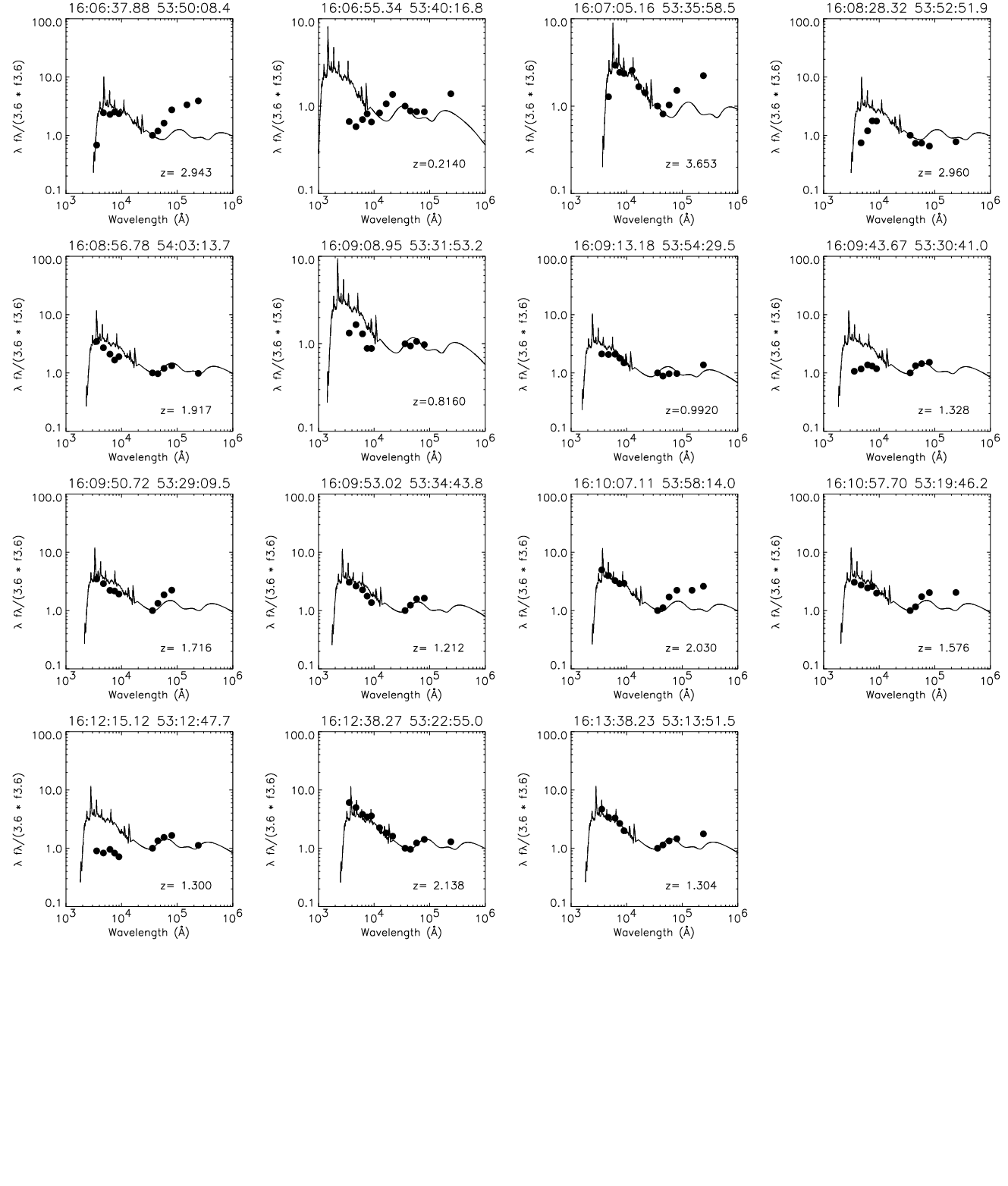}}
\caption{Observed SED for the SWIRE-SDSS quasars (continued).}
\label{figseds2}
\end{figure*}

The Spitzer data in SWIRE EN1 fill a large wavelength gap in the SEDs of AGN. The quasar sample discussed here is optically selected 
and complete down to $i=19.1$. Therefore, when considering their SEDs, one does not see the behavior of an IR selected sample but 
rather the IR properties of optically selected objects with MIR counterparts. Figs. \ref{figseds1} and \ref{figseds2} show the observed 
SEDs (with 2MASS and ELAIS 15 \mums measurements included whenever available; see Section \ref{notes} for details) in $\lambda F_{\lambda}$ (erg/sec/cm$^2$)
overplotted on a template constructed from optically selected (from the Palomar-Green (PG) Survey; hereafter quasar template) type-I AGN 
(from \citealt{polletta04}). Observations and templates have been normalised at 3.6 \mums (or to the $r$-band in the few cases where 
the 3.6 \mums flux was not available). The quasar template was built by extending the composite SDSS quasar optical spectrum \citep{vanden01} 
into the IR with the average SED of 14 PG quasars. At $\lambda<0.7$ \mums the template is the SDSS composite quasar spectrum; the 
averaged PG template is used only at $\lambda>0.7$ \mum. 

The vast majority of the objects have a concave SED, with a change of slope at around 1 \mums in the restframe. The few 
exceptions will be discussed in detail in Section \ref{notes}. The general shape of the SEDs roughly follows the quasar 
template but the use of all 35 quasars provides better sampling, and allows for the refinement of the template, especially
in the IR.

\begin{figure*}
\centerline{
\psfig{figure=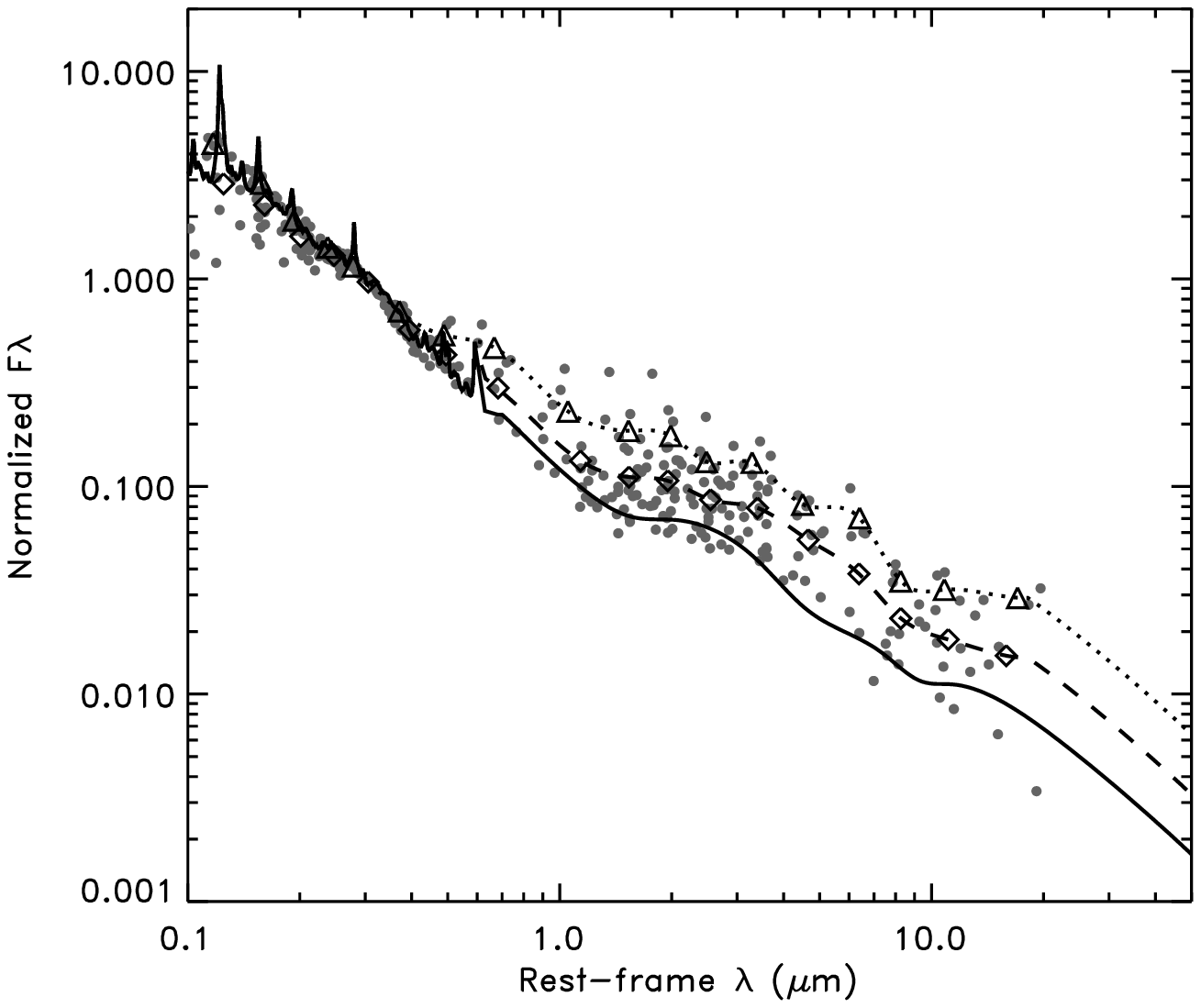}}
\caption{Average (dashed line) and higher MIR flux (dotted line) templates derived from the
SWIRE-SDSS quasars. Diamonds and triangles show averaged data points
in various wavelength bins. The solid line is a template derived from the PG sample
making use of the SDSS composite spectrum in the optical and is showed
here for comparison. The grey circles are the observed data points.}
\label{fignewqsosed}
\end{figure*}

Using the 35 SWIRE-SDSS quasars, two new templates were constructed. The first one (dashed line in Fig. \ref{fignewqsosed}) 
is the average of all of the data, normalised at 0.3 \mums in 18 wavelength bins between 0.1 and 23 \mums (diamonds). The 
upper template (dotted line) is the average of the data points taken from the 25\% objects with the 
highest infrared emission in the same wavelength bins (triangles). The number of flux values in each bin varies from 7 to 37 
with an average of 20 when all sources are used, and from 3 to 10 with an average of six in the case of the second template.
In the optical, at $\lambda<0.6$ \mums for the first template and $\lambda<0.3$ \mums for the second, the more detailed 
quasar template was used since it fits well the average data of both templates. At $\lambda>20$ \mum, since no data are
available, we use the quasar template after scaling it to match the two new templates. The solid curve is the original quasar 
template shown in Figs. \ref{figseds1} and \ref{figseds2}, shown here for comparison. A detailed study of the individual SEDs, 
tori model template fitting and parameter analysis is outside the scope of this work and will be presented in a separate paper.
The two new IR quasar templates defined above are shown in Tables \ref{tabsed1} and \ref{tabsed2}.
\footnote{The SEDs can be downloaded also from the url http://www.iac.es/proyect/swire}

\begin{figure*}[h!]
\centerline{
\psfig{figure=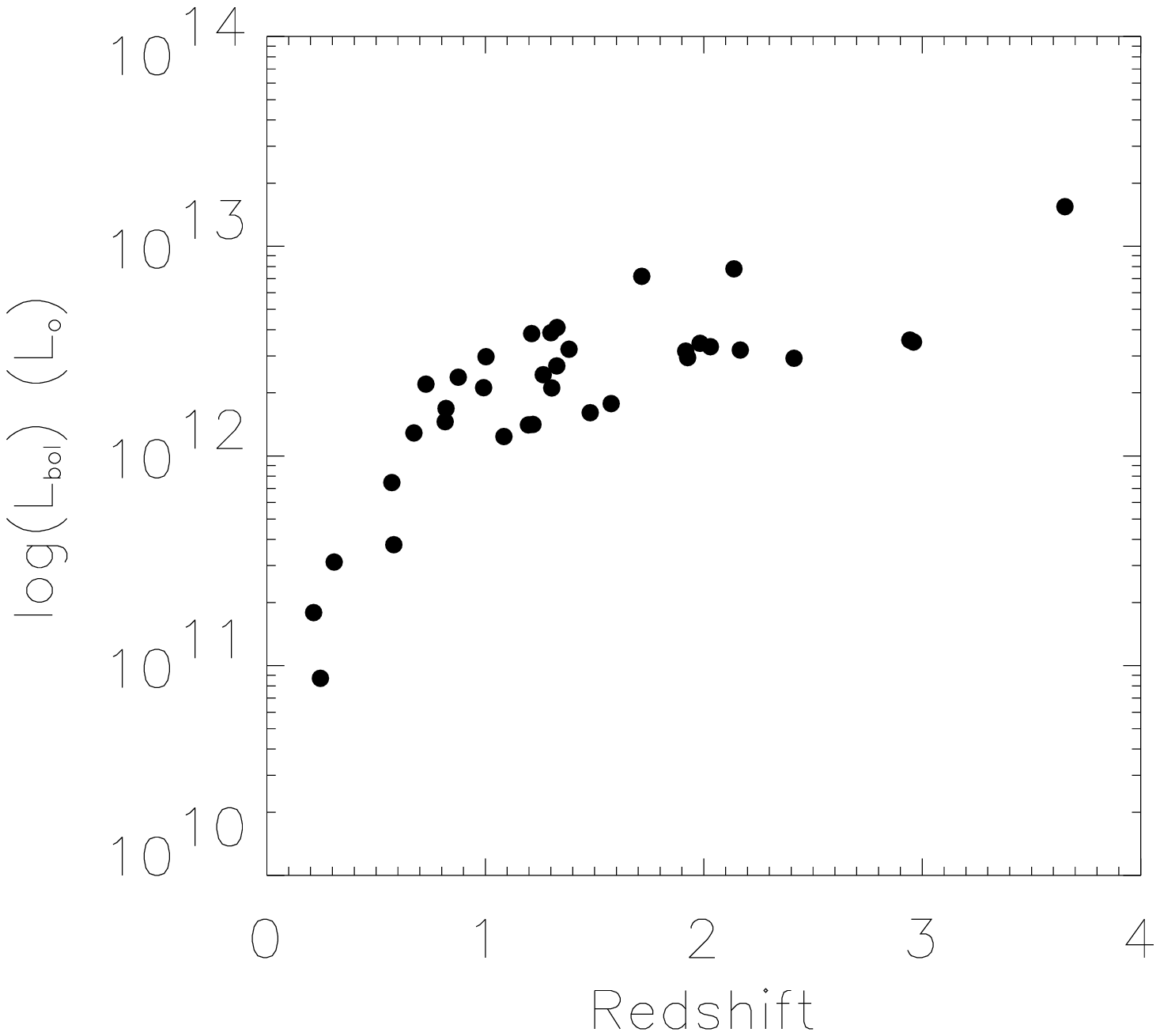}}
\caption{Bolometric luminosities, integrated from $\sim$800 \AA \, to 24 \mum, 
as a function of redshift expressed in units of solar luminosity, computed assuming the
new SEDs.} 
\label{figlbol}
\end{figure*}

Fig. \ref{figlbol} presents the bolometric luminosities (computed assuming an 
($\Omega$, $\Lambda$, H$_0$) = (0.3, 0.7, 70) cosmology), integrated from $\sim$800 \AA \, 
to 24 \mum, as a function of redshift, revealing 85\% of them to be ultraluminous (with L$_{bol} 
\ge 10^{12}$ L$_{\odot}$)  and one hyperluminous (L$_{bol} \ge 10^{13}$ L$_{\odot}$).
The bolometric luminosities have been computed assuming the newly derived SEDs.
The Scott effect clearly dominates this 
flux-limited sample as more distant objects need to be intrinsically brighter in order 
to be part of it.

\section{Black Hole Masses} \label{bhmass}
In this section we investigate a possible correlation between bolometric luminosity and BH mass, as the relationship between these 
two quantities is a measure of the Eddington ratio. Under the assumption that the dynamics of the Broad Line Region (BLR) are dominated 
by the gravity of the central supermassive BH, the latter can be estimated from the relation $M_{BH}\simeq R_{BLR}V^2/G$, where
$R_{BLR}$ is the radius of the BLR and $V$ is the velocity of the line-emitting gas. Traditionally, $V$ is estimated from the full 
width at half maximum (FWHM) of the H$_{\beta}$ line in emission \citep{kaspi00}. However, for quasars with redshifts greater than 
$z \sim 0.8$, H$_{\beta}$ is no longer seen in observed-frame optical spectra, and Mg II has been suggested as an alternative 
estimator \citep{mclure02}.

Here the BH mass is computed as suggested by \cite{mclure04}:
\begin{equation}
\frac{M_{bh}}{M_{\odot}} = 3.2 \left(\frac{\lambda L_{3000}}{10^{37}W}\right)^{0.62} \left(\frac{FWHM(MgII)}{km/sec
}\right)^2
\label{eqMg}
\end{equation}
and
\begin{equation}
\frac{M_{bh}}{M_{\odot}} = 4.7 \left(\frac{\lambda L_{5100}}{10^{37}W}\right)^{0.61} \left(\frac{FWHM(H_{\beta})}{k
m/sec}\right)^2
\label{eqHb}
\end{equation}

\begin{figure*}[h!]
\centerline{
\psfig{figure=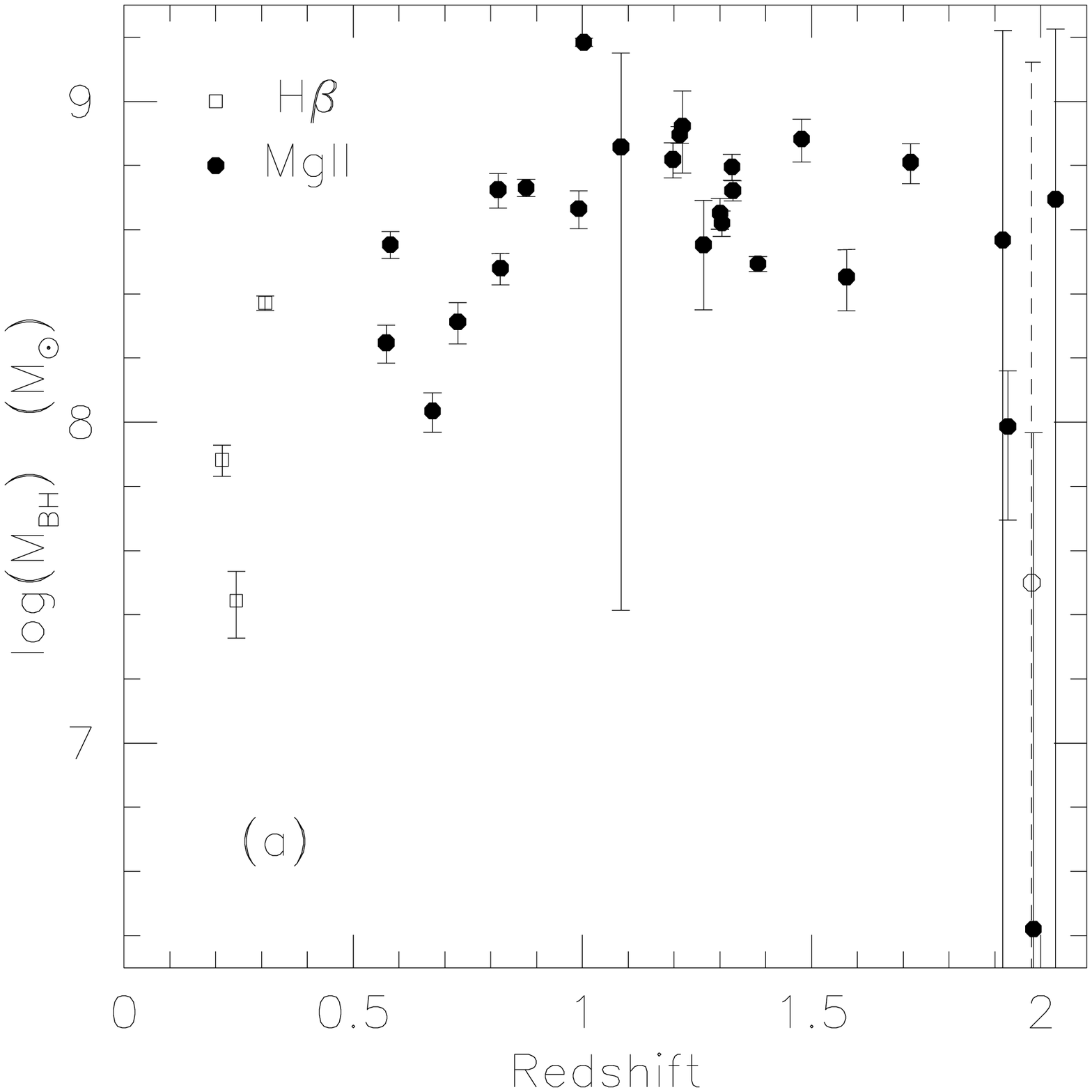,width=8.5cm}
\psfig{figure=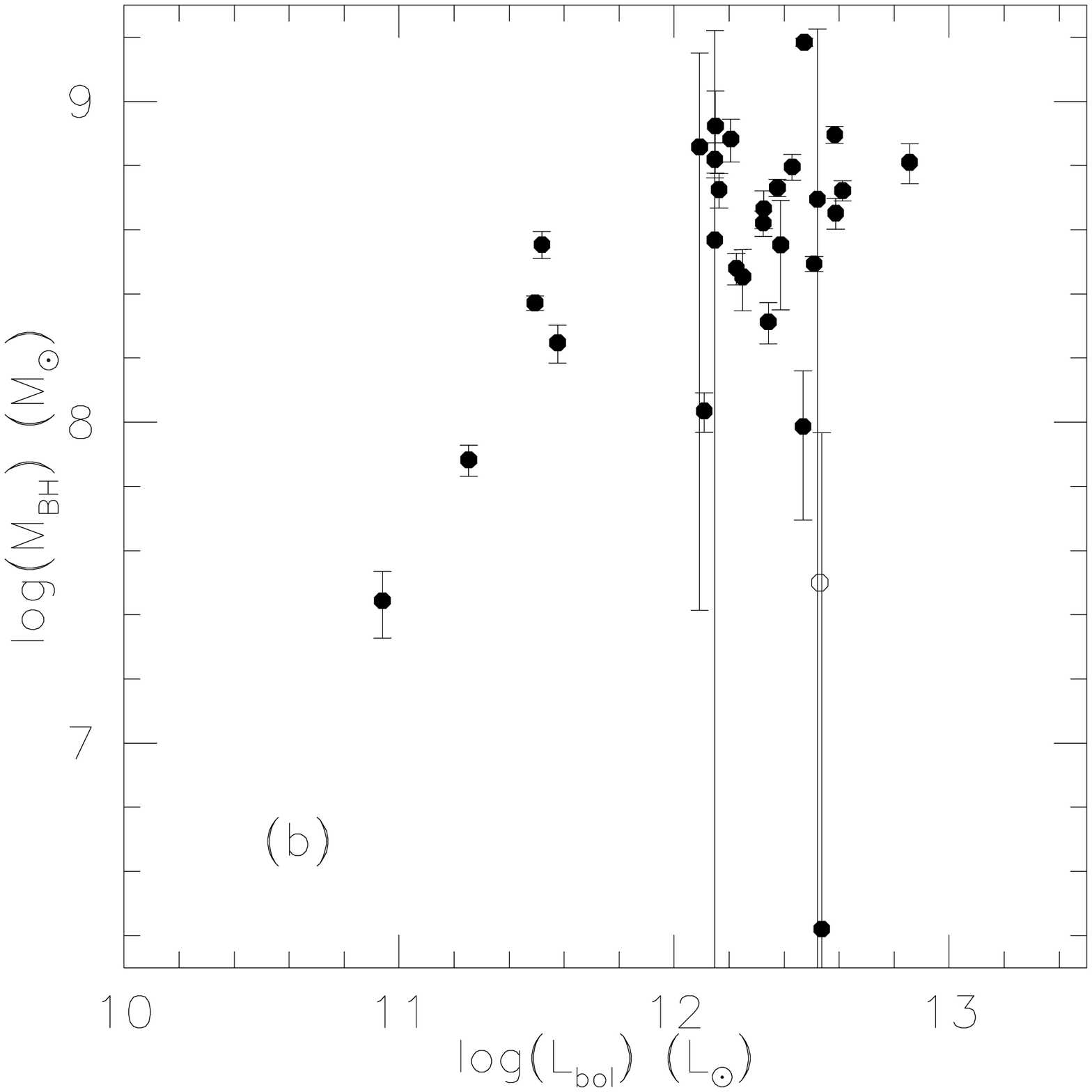,width=8.5cm}}
\caption{(a) BH masses versus redshift. In open squares (filled circles) objects for which H$_{\beta}$ (MgII) 
was used for the estimation. (b) BH mass as a function of bolometric luminosity.
The open circle indicates the value of the BH mass of the dubious object as measured by hand.}
\label{figbhmass}
\end{figure*}

Since the redshift range in which MgII can be observed with optical telescopes is large we chose to use the relation between MgII and 
the (monochromatic) continuum luminosity at 3000 \AA \, (Eq. \ref{eqMg}) whenever available and only use H$_{\beta}$ (Eq. \ref{eqHb}) 
in the few cases when MgII is not present. We make use of the values of the emission lines as measured by the SDSS pipeline and limit 
the study to objects with redshifts lower than, typically, $z=2.1$ as at higher redshift both estimators fall outside the observed spectral 
range. Fig. \ref{figbhmass}(a) shows the BH mass distribution as a function of redshift, with objects for which MgII (H$_{\beta}$) was 
used for the estimation plotted in filled circles (open squares). The error bars are estimated from the uncertainty of the FWHM of the lines. Black hole 
masses tend to increase with redshift up to $z \sim 1$ but then they stabilise around a value of 5 $\times 10^8$ M$_{\odot}$. The very 
low BH mass value of the object at $z \sim 2$ comes from the apparently underestimated value of the MgII line, due to a very noisy 
spectrum. An attempt to measure this line manually gave a value three times larger for the Mg II line, increasing the value of the BH 
mass by roughly an order of magnitude (open symbol in Fig. \ref{figbhmass}). For the rest of the objects, the manual measurements of the 
emission lines agree within 2$\sigma$ with the ones coming from the SDSS pipeline. For the six objects lying in the redshift range from 
0.5 to 0.83 for which both MgII and H$_{\beta}$ were available, BH masses were computed using both estimators. For four of the objects 
the difference in the masses were less than a factor of 1.3, for the other two, however, the estimated masses were almost one order 
of magnitude apart, with H$_{\beta}$ giving the lowest value in both cases.

Fig. \ref{figbhmass}(b) shows the evolution of the bolometric luminosity as a function of BH mass. A trend is seen up to a bolometric luminosity 
of $\sim 10^{12}$ L$_{\odot}$ but it is of low significance, with a Spearman correlation coefficient of 0.32. This is due to small number 
statistics. Note that the determination of BH mass, dependent as it is on optical luminosity, can only be applied to unobscured sources.

\section{Notes on individual Objects}
\label{notes}

Four of the quasars have radio detections: source 18 at
1.4 and 4.85 GHz ($\sim 80$ and 171 mJy, respectively; \citealt{gregory91}; \citealt{white92});
and sources
4, 19 and 27 with FIRST 20cm detections and fluxes 4.20 $\pm$ 0.15
166.23 $\pm$ 0.146 mJy and 43.82 $\pm$ 0.148 mJy, respectively.
Furthermore, two of the radio sources,
18 and 19, also have X-ray counterparts from the ROSAT All Sky
Survey. Finally, six objects have ISO 15 \mums fluxes and six have 2MASS $J$, $H$, $K$
counterparts, as shown in Table \ref{tabancdata}, along with the radion information.

\begin{figure*}[h!]
\centerline{
\psfig{figure=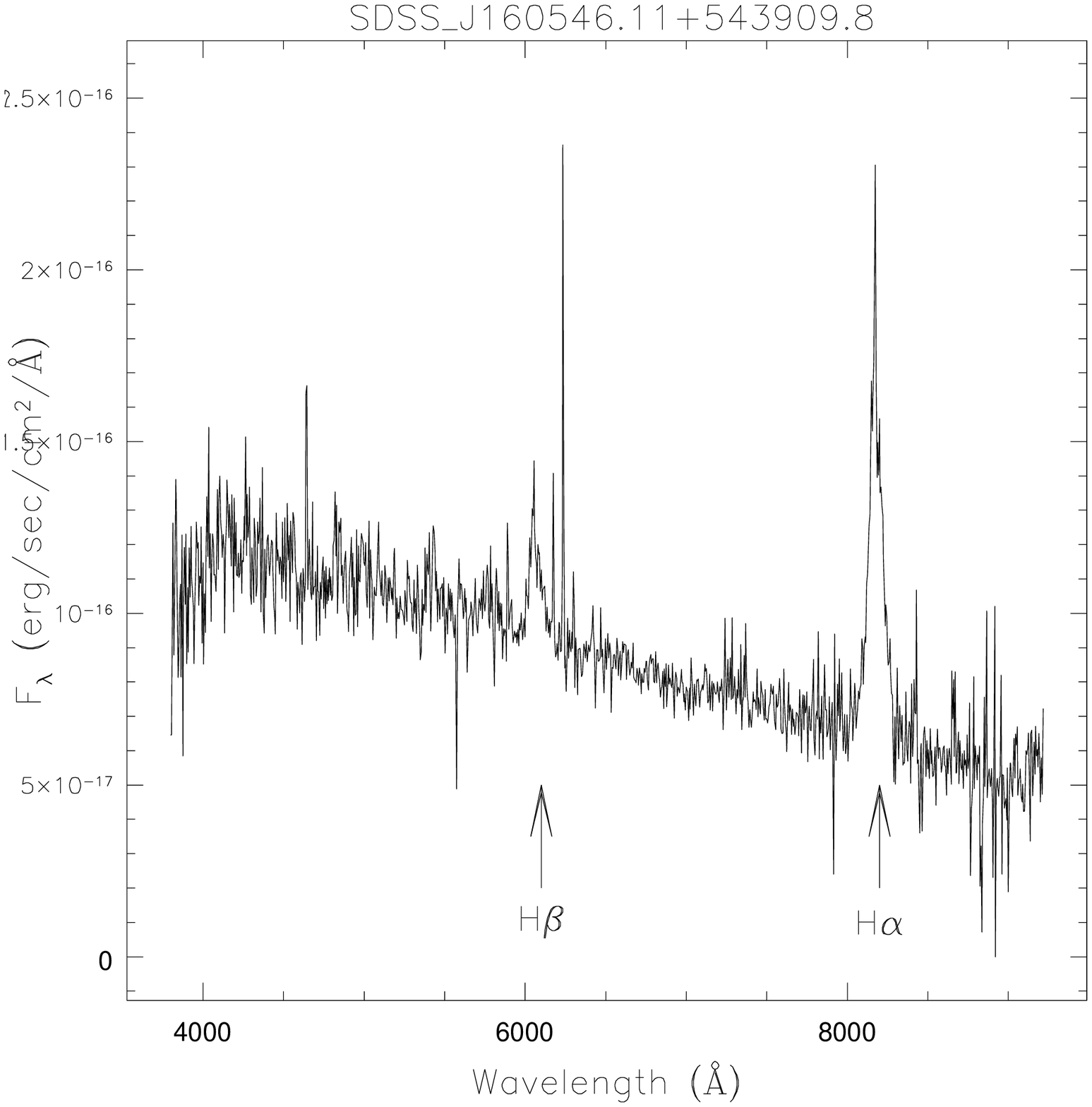,width=8.5cm}
\psfig{figure=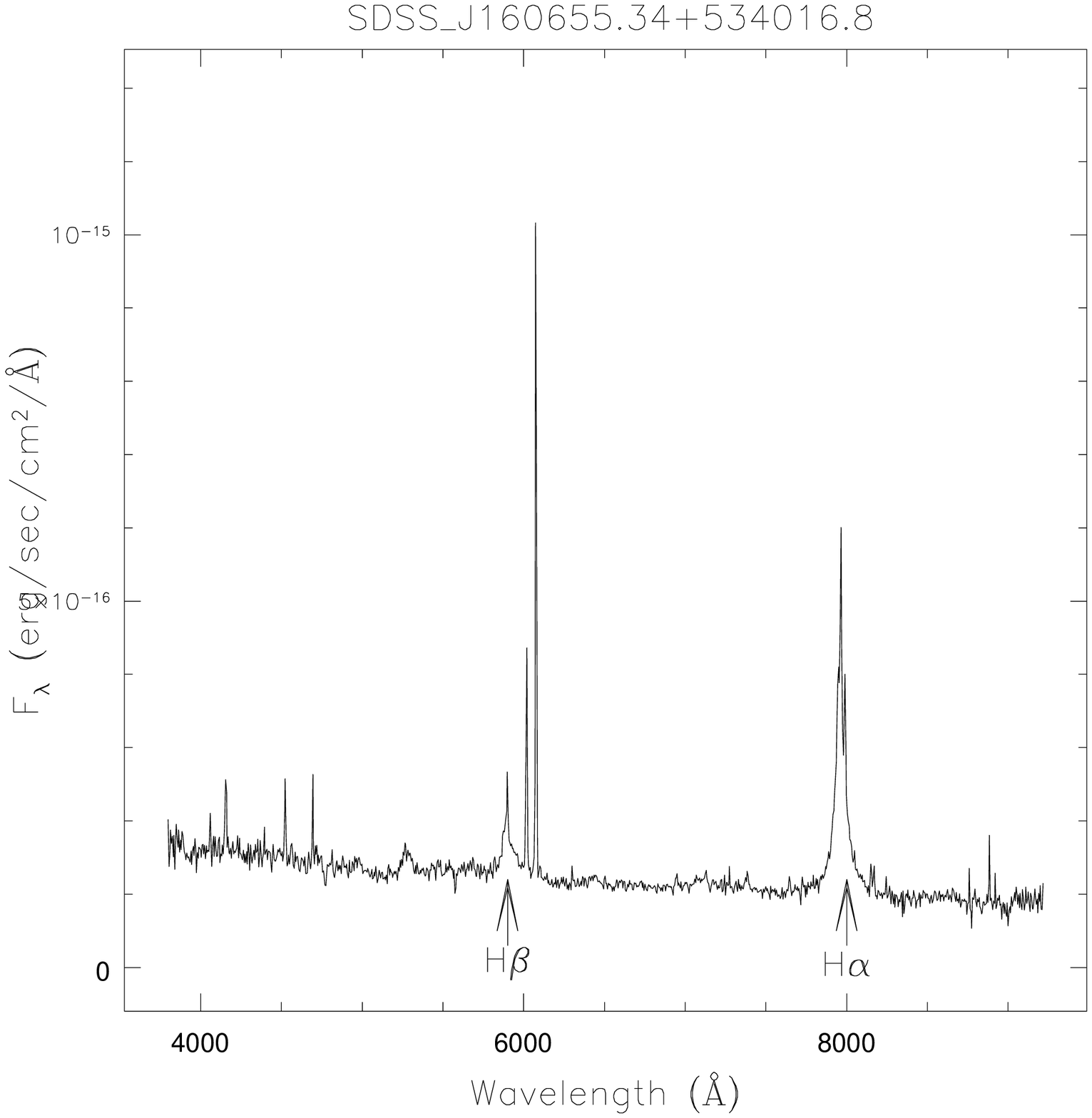,width=8.5cm}}
\centerline{
\psfig{figure=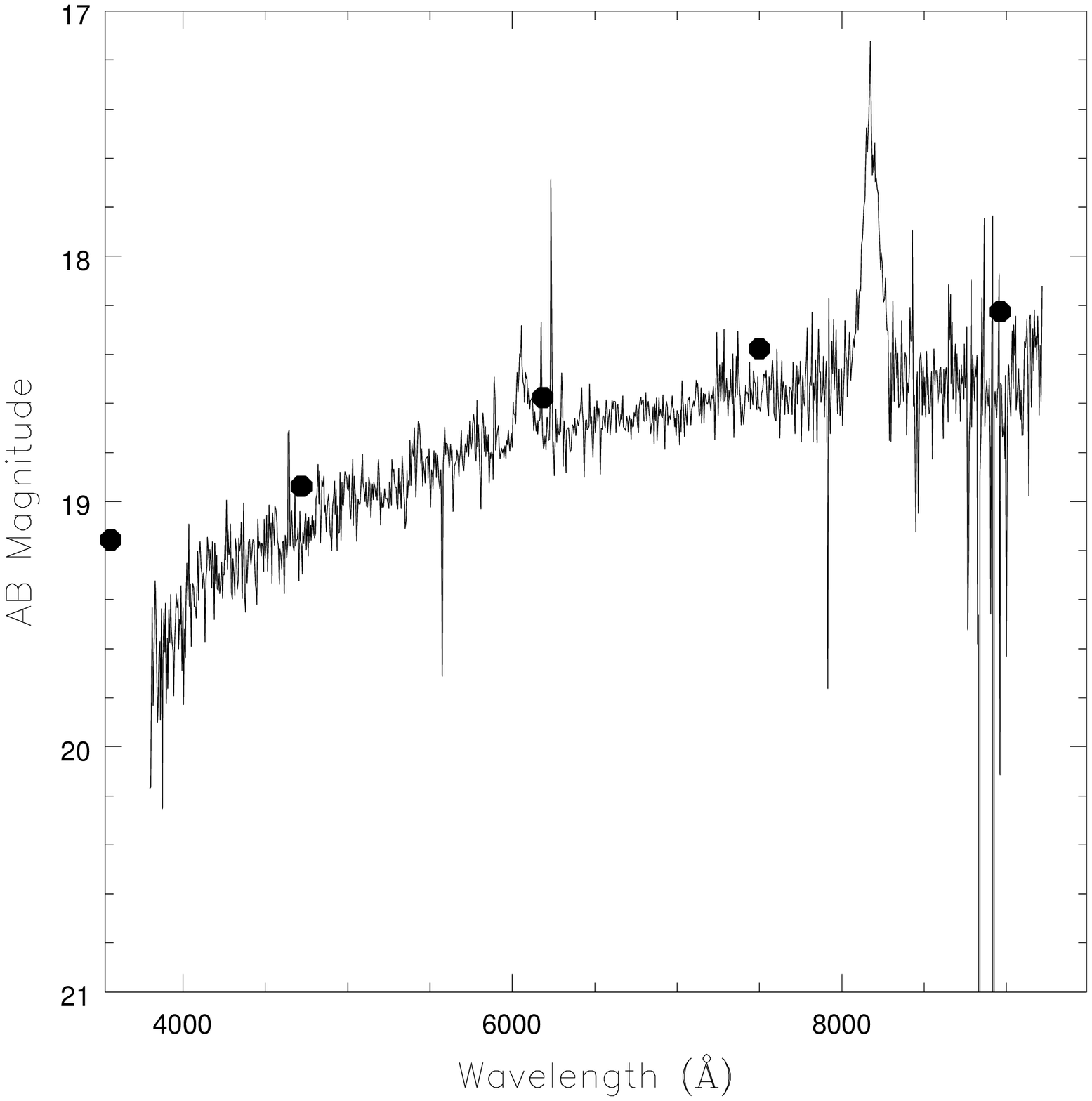,width=8.5cm}
\psfig{figure=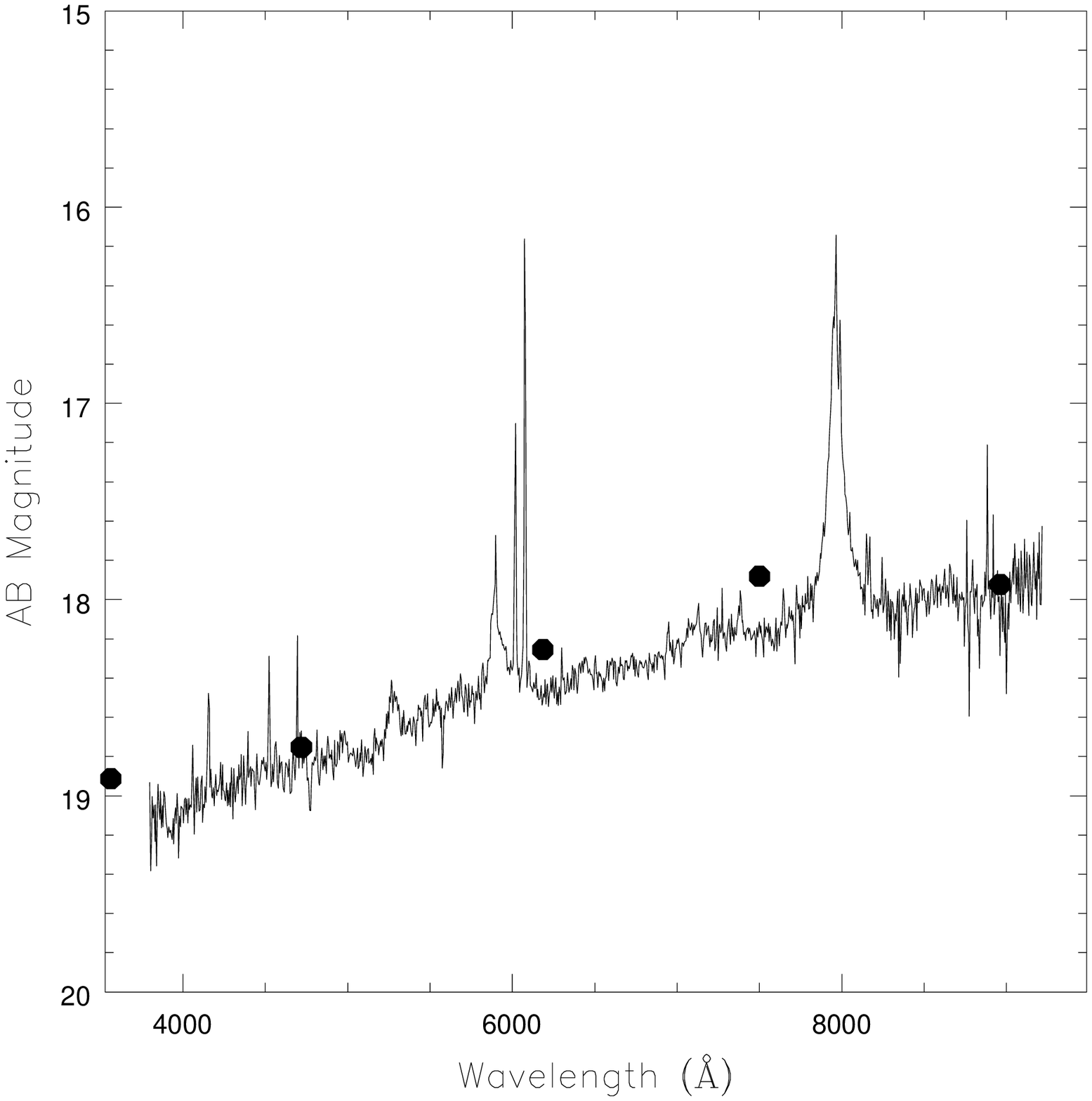,width=8.5cm}}
\caption{Upper panels: SDSS spectrum of objects 18 and 22
showing broad emission lines, characteristic of a type-1 quasar. Object 22
also presents strong narrow
forbidden lines, as in type-II AGN. Lower panels: spectra and magnitudes of the two 
objects are shown in the AB system, demonstrating signs of variability.}
\label{figspec}
\end{figure*}

Object 18, at a redshift of 0.245, is particularly interesting. As can be seen in Fig. \ref{figseds1} (second object in the last row), its SED does 
not have the concave shape that the majority of the objects have. Its SED is instead convex, and can be fitted by two power laws, a relatively flat 
one in the optical and a much steeper one in the IR. Even though this object is a type-I quasar with broad emission lines, as can be seen in the 
left panel of Fig. \ref{figspec}, the behavior of its SED is that of a blazar. Its inverted radio spectrum (higher flux in higher frequencies) 
also points towards this conclusion.

A similar behavior is observed for source 22 
at $z=0.214$, whose spectrum is shown in the right column of Fig. \ref{figspec}. Unfortunately no other data are available for this object.
Both objects show signs of variability, an important characteristic of all blazars, between the times the spectroscopy and the photometry were made,
as can be seen in the lower panels of Fig. \ref{figspec}, where the spectra and magnitudes of the two objects are shown in the AB system.
For a comprehensive study of quasar variability based on this type of comparison see \cite{vanden04}. The photometric errors are of the size of 
the filled circles (or smaller) and can not be seen on the plot.

Finally, sources 21 and 24,
whose optical spectra are considerably redder than the quasar template (Fig. \ref{figseds2}), are in fact broad absorption line quasars, as visual
inspection of its spectrum revealed. Recent studies relate reddened quasars with the presence of 
broad absorpion lines in their spectra (e.g. \citealt{reichard03} 
and references therein).

\section{Discussion} \label{discuss}
This paper presents an analysis of a sample of 35 spectroscopically confirmed SDSS quasars within the SWIRE EN1 field, observed with Spitzer as part of the
SWIRE Legacy Survey. Even though SWIRE EN1 is a small part of the whole 50 deg$^2$ SWIRE survey ($\sim$18\%) and SDSS DR2 only covers a third of it,
this study gives a good initial indication of the AGN science that can be carried out with SWIRE data.

All SDSS quasars within the SWIRE EN1 field have Spitzer counterparts. The distinct location of the spectroscopically confirmed quasars on 
MIR color-color plots greatly contributes to the selection of type-I AGN candidates, independent of their optical properties. They are redder than the 
majority of the population at least up to 8 \mums (restframe from 1.7 to 6.1 \mums for the redshift range covered) and continue forming a rather compact 
clump in the redder colors.
Appropriate color-color combinations can be found in order to separate quasars of specific (but large) redshift ranges from the various types of contaminant 
galaxies. A comparison between model predictions and selected quasar candidates indicates the existance of a large number of obscured AGN, some of which
probably reside among the proposed candidates.

BH masses, estimated from emission lines, show a tendency with redshift up to $z \sim 1$, however the small number statistics
do not allow any conclusion on the dependence or not on bolometric luminosity.

Even though large multi-wavelength spectroscopic efforts are still required in order to understand the nature and propeties of AGN, we demonstrate that
Spitzer MIR data contribute significantly towards this goal.

\acknowledgments
This work is based on observations made with the {\it Spitzer Space Telescope},
which is operated by the Jet Propulsion Laboratory, California Institute of
Technology under NASA contract 1407.
Support for this work, part of the Spitzer Space Telescope Legacy Science
Program, was provided by NASA through an award issued by the Jet Propulsion
Laboratory, California Institute of Technology under NASA contract 1407.

Funding for the creation and distribution of the SDSS Archive has been provided 
by the Alfred P. Sloan Foundation, the Participating Institutions, the National 
Aeronautics and Space Administration, the National Science Foundation, the U.S. 
Department of Energy, the Japanese Monbukagakusho, and the Max Planck Society. 
The SDSS Web site is http://www.sdss.org/.
The SDSS is managed by the Astrophysical Research Consortium (ARC) for the 
Participating Institutions. The Participating Institutions are The University of 
Chicago, Fermilab, the Institute for Advanced Study, the Japan Participation Group, 
The Johns Hopkins University, Los Alamos National Laboratory, the 
Max-Planck-Institute for Astronomy (MPIA), the Max-Planck-Institute for 
Astrophysics (MPA), New Mexico State University, University of Pittsburgh, 
Princeton University, the United States Naval Observatory, and the University of 
Washington.

This research makes use of the NASA/IPAC Extragalactic Database (NED) which 
is operated by the Jet Propulsion Laboratory, California Institute of Technology, 
under contract with the National Aeronautics and Space Administration.
It also makes use of data products from the Two Micron All Sky Survey, 
which is a joint project of the University of Massachusetts and the Infrared 
Processing and Analysis Center/California Institute of Technology, funded by the 
National Aeronautics and Space Administration and the National Science Foundation.
This work was supported in part by the Spanish Ministerio de
Ciencia y Tecnologia (Grants Nr. PB1998-0409-C02-01 and ESP2002-03716)
and by the EC network "POE'' (Grant Nr. HPRN-CT-2000-00138).

This work also makes use of
data products provided by the CASU INT Wide Field Survey.
The INT and WHT telescopes are operated on the
island of La Palma by the Isaac Newton Group in the Spanish Observatorio
del Roque de los Muchachos of the Instituto de Astrofisica de Canarias.

Thanks are due to J. Fritz for very useful discussions.
We would also like to thank the referee, Michael Strauss, for his very 
constructive comments that greatly improved the presentation of our work.

\clearpage

\newpage

\begin{deluxetable}{lccccccccccccc}
\rotate
\tablecolumns{14}
\setlength{\tabcolsep}{0.03in}
\tablewidth{0pt}
\tablecaption{Positions, redshift, optical magnitudes and aperture fluxes for the 35 spectroscopically confirmed
quasars. The SDSS magnitudes are in the AB system, fluxes are given in mJy. Note that
the errors in the IRAC1 and IRAC2 fluxes are lower than 0.01 mJy (typically of the order 
of a few \textmu Jy) and are not reported here.}
\tablehead{
\colhead{Seq Nr} & \colhead{RA} & \colhead{Dec} & \colhead{z} & \colhead{$u$} & \colhead{$g$}
& \colhead{$r$}& \colhead{$i$} & \colhead{$z$} & \colhead{\it IRAC1} & \colhead{\it IRAC2} 
& \colhead{\it IRAC3} & \colhead{\it IRAC4} & \colhead{\it MIPS24} }
\startdata
1$^b$&15:56:32.53&54:43:35.9&1.003&18.34&18.23&18.05&18.21&18.25       &  --  &  -- &  --	  --  &  --	    --  & 9.51 $\pm$ 0.03 \\
2&15:57:26.63&55:07:40.9&1.383&19.01&18.88&18.80&18.79&18.78           & 0.42 &  -- & 1.03 $\pm$ 0.01 &  --	    --  &  --	      --  \\
3$^b$&15:58:54.96&54:28:36.2&1.085&19.31&19.26&19.08&19.22&19.25       &  --  &  -- &  --	  --  &  --	    --  & 1.04 $\pm$ 0.02 \\
4$^{d,e}$&15:59:36.13&54:42:03.8&0.308&18.60&18.46&18.29&18.40&17.89   & 1.16 & 1.44& 2.11 $\pm$ 0.01 & 3.27 $\pm$ 0.01 & 13.7 $\pm$ 0.02 \\
5&16:00:04.33&55:04:29.9&1.983&18.97&18.85&18.81&18.78&18.70           & 0.20 & 0.33& 0.62 $\pm$ 0.01 & 1.02 $\pm$ 0.01 & 3.27 $\pm$ 0.02 \\
6&16:00:15.68&55:22:59.9&0.673&18.96&18.56&18.55&18.42&18.35           & 0.88 & 1.16& 1.48 $\pm$ 0.01 & 2.01 $\pm$ 0.01 & 4.29 $\pm$ 0.02 \\
7&16:01:00.07&55:11:04.8&1.218&19.44&19.45&19.13&19.21&19.38           & 0.25 & 0.34& 0.45 $\pm$ 0.01 & 0.57 $\pm$ 0.01 & 1.37 $\pm$ 0.02 \\
8&16:01:28.54&54:45:21.3&0.728&18.42&18.16&18.09&18.07&18.02           & 1.28 & 1.88& 2.77 $\pm$ 0.02 & 3.83 $\pm$ 0.01 & 12.1 $\pm$ 0.02 \\
9$^b$&16:01:39.49&54:11:13.1&2.167&19.58&19.30&19.28&19.18&18.95       &  --  & 0.16&  --	  --  & 0.48 $\pm$ 0.01 & 1.07 $\pm$ 0.02 \\
10&16:02:30.02&54:36:58.5&1.327&18.92&18.70&18.42&18.42&18.51          & 0.39 & 0.63& 1.08 $\pm$ 0.01 & 1.46 $\pm$ 0.01 & 4.59 $\pm$ 0.02 \\
11&16:02:44.29&54:45:17.8&1.480&19.38&19.30&19.20&19.10&19.08          & 0.18 & 0.26& 0.40 $\pm$ 0.01 & 0.58 $\pm$ 0.01 & 1.80 $\pm$ 0.02 \\
12$^c$&16:02:50.95&54:50:57.9&1.197&19.69&19.56&19.22&19.19&19.20      & 0.27 & 0.41& 0.68 $\pm$ 0.01 & 1.06 $\pm$ 0.01 & 3.80 $\pm$ 0.02 \\
13&16:03:01.86&54:55:43.6&1.926&19.70&19.26&19.11&18.78&18.54          & 0.19 & 0.25& 0.40 $\pm$ 0.01 & 0.59 $\pm$ 0.01 & 1.39 $\pm$ 0.02 \\
14&16:03:41.44&54:15:01.5&0.581&19.63&19.29&19.32&19.20&19.15          & 0.35 & 0.39& 0.50 $\pm$ 0.01 & 0.58 $\pm$ 0.01 & 1.13 $\pm$ 0.02 \\
15&16:04:29.01&54:43:33.9&1.265&19.17&18.79&18.41&18.38&18.40          & 0.39 & 0.54& 0.76 $\pm$ 0.01 & 1.06 $\pm$ 0.01 & 1.89 $\pm$ 0.02 \\
16&16:04:50.72&54:15:40.4&2.413&21.16&20.20&20.06&19.74&19.22          & 0.13 & 0.13& 0.20 $\pm$ 0.01 & 0.30 $\pm$ 0.01 & 0.49 $\pm$ 0.02 \\
17$^{c,g}$&16:05:23.10&54:56:13.3&0.572&19.17&18.86&18.92&18.77&18.82  & 0.72 & 0.96& 1.43 $\pm$ 0.01 & 1.76 $\pm$ 0.01 & 4.49 $\pm$ 0.02 \\
18$^{d,e,g}$&16:05:46.11&54:39:09.8&0.245&19.16&18.94&18.58&18.38&18.23& 0.54 & 0.55& 0.67 $\pm$ 0.01 & 0.73 $\pm$ 0.01 & 1.13 $\pm$ 0.02 \\
19$^{c,d,e}$&16:06:23.56&54:05:55.8&0.876&17.95&17.68&17.62&17.68&17.65& 0.92 & 1.22& 1.79 $\pm$ 0.01 & 2.38 $\pm$ 0.01 & 7.33 $\pm$ 0.02 \\
20$^c$&16:06:30.60&54:20:07.5&0.820&19.03&18.77&18.73&18.83&18.61      & 0.76 & 1.07& 1.59 $\pm$ 0.01 & 2.21 $\pm$ 0.01 & 5.23 $\pm$ 0.02 \\
21$^c$&16:06:37.88&53:50:08.4&2.943&21.68&19.99&19.76&19.45&19.33      & 0.12 & 0.17& 0.30 $\pm$ 0.01 & 0.70 $\pm$ 0.01 & 2.99 $\pm$ 0.02 \\
22$^d$&16:06:55.34&53:40:16.8&0.214&18.91&18.75&18.26&17.88&17.92      & 1.50 & 1.65& 2.09 $\pm$ 0.01 & 2.87 $\pm$ 0.01 & 13.8 $\pm$ 0.02 \\
23$^d$&16:07:05.16&53:35:58.5&3.653&23.61&19.35&18.15&18.14&17.98      & 0.40 & 0.40& 0.66 $\pm$ 0.01 & 1.33 $\pm$ 0.01 & 5.89 $\pm$ 0.02 \\
24&16:08:28.32&53:52:51.9&2.960&25.83&21.25&20.44&19.80&19.62          & 0.12 & 0.11& 0.14 $\pm$ 0.01 & 0.17 $\pm$ 0.01 & 0.61 $\pm$ 0.02 \\
25&16:08:56.78&54:03:13.7&1.917&19.29&19.25&19.23&19.28&18.94          & 0.20 & 0.25& 0.39 $\pm$ 0.01 & 0.60 $\pm$ 0.01 & 1.34 $\pm$ 0.02 \\
26$^a$&16:09:08.95&53:31:53.2&0.816&19.05&18.50&18.48&18.68&18.49      & 0.66 & 0.78& 1.13 $\pm$ 0.01 & 1.44 $\pm$ 0.01 &  --	      --  \\
27$^e$&16:09:13.18&53:54:29.5&0.992&18.63&18.34&18.04&18.00&18.01      & 0.62 & 0.68& 0.96 $\pm$ 0.01 & 1.34 $\pm$ 0.01 & 5.65 $\pm$ 0.03 \\
28$^a$&16:09:43.67&53:30:41.0&1.328&19.42&19.01&18.54&18.39&18.30      & 0.59 & 0.98& 1.37 $\pm$ 0.01 & 2.00 $\pm$ 0.01 &  --	      --  \\
29$^a$&16:09:50.72&53:29:09.5&1.716&18.15&18.04&18.04&17.86&17.79      & 0.58 & 0.97& 1.74 $\pm$ 0.01 & 2.89 $\pm$ 0.01 &  --	      --  \\
30$^{a,f}$&16:09:53.02&53:34:43.8&1.212&18.10&17.96&17.83&17.89&17.97  & 0.69 & 1.06& 1.75 $\pm$ 0.01 & 2.49 $\pm$ 0.01 &  --	      --  \\
31$^c$&16:10:07.11&53:58:14.0&2.030&18.99&18.92&18.85&18.77&18.56      & 0.19 & 0.26& 0.52 $\pm$ 0.01 & 0.93 $\pm$ 0.01 & 3.26 $\pm$ 0.03 \\
32&16:10:57.70&53:19:46.2&1.576&19.61&19.42&19.25&18.99&19.07          & 0.17 & 0.25& 0.48 $\pm$ 0.01 & 0.78 $\pm$ 0.01 & 2.35 $\pm$ 0.03 \\
33&16:12:15.12&53:12:47.7&1.300&19.61&19.39&18.95&18.89&18.86          & 0.59 & 0.98& 1.45 $\pm$ 0.01 & 2.16 $\pm$ 0.01 & 4.38 $\pm$ 0.02 \\
34$^d$&16:12:38.27&53:22:55.0&2.138&17.92&17.82&17.82&17.72&17.48      & 0.41 & 0.49& 0.82 $\pm$ 0.01 & 1.30 $\pm$ 0.01 & 3.56 $\pm$ 0.02 \\
35&16:13:38.23&53:13:51.5&1.304&18.49&18.53&18.27&18.29&18.41          & 0.32 & 0.45& 0.68 $\pm$ 0.01 & 1.03 $\pm$ 0.01 & 3.72 $\pm$ 0.02 \\
\label{tabquasars}\enddata
\tablenotetext{a} {Object in the MIPS 24 \mums gap}
\tablenotetext{b} {Object at the border of the field not (or partially) covered by IRAC}
\tablenotetext{c} {Object with ELAIS 15 \mums detection}
\tablenotetext{d} {Object with 2MASS $JHK$ counterparts}
\tablenotetext{e} {Object with radio detection}
\tablenotetext{f} {Object with IRAS detections}
\tablenotetext{g} {Object with RASS detections}
\tablenotetext{c,d,e} {2MASS magnitudes, ISO 15 \mums and radio fluxes can be found 
in Table \ref{tabancdata}}
\end{deluxetable}

\begin{deluxetable}{cccccccccc}
\tablecolumns{10}
\setlength{\tabcolsep}{0.04in}
\tablewidth{0pt}
\tablecaption{Quasar SED template created using the 35 SWIRE-SDSS quasars.
spanning the wavelength range between $\sim 800$ and $2 \times 10^6$ \AA. 
The fluxes are normalised at $\lambda=0.3$ \textmu m. The complete version of this table is in the 
electronic edition of the Journal. The printed edition contains only a sample.}
\tablehead{
\colhead{$\lambda$ (\textmu m)}& \colhead{f$_{\lambda}$/f$_{0.3 \mu}$} &
\colhead{$\lambda$ (\textmu m)}& \colhead{f$_{\lambda}$/f$_{0.3 \mu}$} &
\colhead{$\lambda$ (\textmu m)}& \colhead{f$_{\lambda}$/f$_{0.3 \mu}$} &
\colhead{$\lambda$ (\textmu m)}& \colhead{f$_{\lambda}$/f$_{0.3 \mu}$} &
\colhead{$\lambda$ (\textmu m)}& \colhead{f$_{\lambda}$/f$_{0.3 \mu}$} }
\startdata
0.079 & 0.369  & 0.195 & 1.904 & 0.311 & 0.966 & 0.428 & 0.489 & 0.544 & 0.293 \\
0.083 & 0.546  & 0.199 & 1.778 & 0.315 & 0.979 & 0.432 & 0.541 & 0.548 & 0.294 \\
0.087 & 1.346  & 0.203 & 1.691 & 0.319 & 0.969 & 0.436 & 0.539 & 0.552 & 0.298 \\
0.091 & 2.197  & 0.207 & 1.741 & 0.323 & 0.944 & 0.440 & 0.481 & 0.556 & 0.309 \\
0.095 & 3.190  & 0.211 & 1.611 & 0.327 & 0.908 & 0.444 & 0.455 & 0.560 & 0.313 \\
0.099 & 3.293  & 0.215 & 1.513 & 0.331 & 0.876 & 0.448 & 0.457 & 0.564 & 0.304 \\
0.103 & 4.694  & 0.219 & 1.446 & 0.335 & 0.852 & 0.452 & 0.466 & 0.568 & 0.290 \\
0.107 & 3.550  & 0.223 & 1.413 & 0.339 & 0.853 & 0.456 & 0.465 & 0.572 & 0.278 \\
0.111 & 3.128  & 0.227 & 1.439 & 0.343 & 0.840 & 0.460 & 0.450 & 0.576 & 0.274 \\
0.115 & 3.006  & 0.231 & 1.495 & 0.347 & 0.807 & 0.464 & 0.440 & 0.580 & 0.284 \\
0.119 & 4.272  & 0.235 & 1.444 & 0.351 & 0.771 & 0.468 & 0.417 & 0.584 & 0.317 \\
0.123 & 7.052  & 0.239 & 1.436 & 0.355 & 0.743 & 0.472 & 0.398 & 0.588 & 0.382 \\
0.127 & 3.706  & 0.243 & 1.476 & 0.359 & 0.729 & 0.476 & 0.391 & 0.843 & 0.210 \\
0.131 & 3.405  & 0.247 & 1.432 & 0.363 & 0.711 & 0.480 & 0.426 & 0.995 & 0.159 \\
0.135 & 3.038  & 0.251 & 1.381 & 0.367 & 0.693 & 0.484 & 0.534 & 1.505 & 0.111 \\
0.139 & 3.644  & 0.255 & 1.324 & 0.372 & 0.690 & 0.488 & 0.534 & 2.276 & 0.094 \\
0.143 & 2.804  & 0.259 & 1.270 & 0.376 & 0.665 & 0.492 & 0.455 & 3.442 & 0.078 \\
0.147 & 2.667  & 0.263 & 1.238 & 0.380 & 0.624 & 0.496 & 0.449 & 5.205 & 0.049 \\
0.151 & 2.910  & 0.267 & 1.197 & 0.384 & 0.620 & 0.500 & 0.467 & 7.871 & 0.025 \\
0.155 & 4.357  & 0.271 & 1.227 & 0.388 & 0.607 & 0.504 & 0.431 & 11.90 & 0.017 \\
0.159 & 2.695  & 0.275 & 1.315 & 0.392 & 0.571 & 0.508 & 0.339 & 18.15 & 0.014 \\
0.163 & 2.667  & 0.279 & 1.874 & 0.396 & 0.552 & 0.512 & 0.339 & 23.14 & 0.010 \\
0.167 & 2.427  & 0.283 & 1.284 & 0.400 & 0.533 & 0.516 & 0.352 & 29.51 & 0.007 \\
0.171 & 2.296  & 0.287 & 1.141 & 0.404 & 0.514 & 0.520 & 0.351 & 37.63 & 0.005 \\
0.175 & 2.245  & 0.291 & 1.067 & 0.408 & 0.531 & 0.524 & 0.348 & 47.98 & 0.003 \\
0.179 & 2.126  & 0.295 & 1.087 & 0.412 & 0.523 & 0.528 & 0.346 & 61.17 & 0.002 \\
0.183 & 2.059  & 0.300 & 1.000 & 0.416 & 0.490 & 0.532 & 0.336 & 77.99 & 0.001 \\
0.187 & 2.235  & 0.303 & 0.940 & 0.420 & 0.477 & 0.536 & 0.315 & --    & --    \\
0.191 & 2.562  & 0.307 & 0.954 & 0.424 & 0.472 & 0.540 & 0.300 & --    & --    \\
\label{tabsed1}
\enddata
\end{deluxetable}

\begin{deluxetable}{cccccccc}
\tablecolumns{8}
\setlength{\tabcolsep}{0.03in}
\tablewidth{0pt}
\tablecaption{Quasar SED template created using the SWIRE-SDSS quasar subsample with the
largest IR fluxes. The complete version of this table is in the electronic edition of
the Journal. The printed edition contains only a sample.}
\tablehead{
\colhead{$\lambda$ (\textmu m)}& \colhead{f$_{\lambda}$/f$_{0.3 \mu}$} &
\colhead{$\lambda$ (\textmu m)}& \colhead{f$_{\lambda}$/f$_{0.3 \mu}$} &
\colhead{$\lambda$ (\textmu m)}& \colhead{f$_{\lambda}$/f$_{0.3 \mu}$} &
\colhead{$\lambda$ (\textmu m)}& \colhead{f$_{\lambda}$/f$_{0.3 \mu}$} }
\startdata
0.079 &  0.369 & 0.148 &  2.674 & 0.216 &  1.514 & 0.285 &  1.204 \\
0.082 &  0.645 & 0.150 &  2.829 & 0.219 &  1.446 & 0.288 &  1.132 \\
0.084 &  0.850 & 0.153 &  4.000 & 0.222 &  1.436 & 0.291 &  1.060 \\
0.087 &  1.418 & 0.156 &  3.251 & 0.225 &  1.419 & 0.294 &  1.098 \\
0.090 &  2.027 & 0.159 &  2.695 & 0.228 &  1.448 & 0.296 &  1.076 \\
0.093 &  2.975 & 0.162 &  2.663 & 0.231 &  1.494 & 0.300 &  1.000 \\
0.096 &  2.931 & 0.165 &  2.540 & 0.233 &  1.487 & 0.417 &  0.601 \\
0.099 &  3.293 & 0.168 &  2.368 & 0.236 &  1.433 & 0.513 &  0.526 \\
0.102 &  3.579 & 0.170 &  2.289 & 0.239 &  1.436 & 0.631 &  0.487 \\
0.105 &  3.455 & 0.173 &  2.275 & 0.242 &  1.505 & 0.776 &  0.380 \\
0.107 &  3.488 & 0.176 &  2.185 & 0.245 &  1.430 & 0.955 &  0.265 \\
0.110 &  3.073 & 0.179 &  2.126 & 0.248 &  1.429 & 1.037 &  0.234 \\
0.113 &  3.018 & 0.182 &  2.059 & 0.251 &  1.395 & 1.929 &  0.180 \\
0.116 &  3.123 & 0.185 &  2.199 & 0.253 &  1.360 & 3.587 &  0.117 \\
0.119 &  4.272 & 0.188 &  2.294 & 0.256 &  1.307 & 6.671 &  0.064 \\
0.122 &  9.076 & 0.190 &  2.717 & 0.259 &  1.270 & 12.40 &  0.031 \\
0.125 &  5.074 & 0.193 &  2.035 & 0.262 &  1.256 & 21.00 &  0.024 \\
0.127 &  3.421 & 0.196 &  1.821 & 0.265 &  1.173 & 30.23 &  0.014 \\
0.130 &  3.525 & 0.199 &  1.778 & 0.268 &  1.185 & 43.53 &  0.008 \\
0.133 &  3.282 & 0.202 &  1.722 & 0.271 &  1.216 & 62.67 &  0.004 \\
0.136 &  3.041 & 0.205 &  1.711 & 0.274 &  1.279 & 90.24 &  0.002 \\
0.139 &  3.644 & 0.208 &  1.735 & 0.276 &  1.345 & 129.9 &  0.001 \\
0.142 &  2.940 & 0.211 &  1.644 & 0.279 &  1.874 & --	 & --	  \\
0.145 &  2.702 & 0.213 &  1.559 & 0.282 &  1.347 & --	 & --	  \\
\label{tabsed2}
\enddata
\end{deluxetable}

\begin{deluxetable}{ccccccccc}
\rotate
\tablecolumns{9}
\setlength{\tabcolsep}{0.03in}
\tablewidth{0pt}
\tablecaption{2MASS Vega magnitudes, ISO fluxes (from \citealt{vaccari04}) radio fluxes
and related errors (whenever available)
for the quasars with near-IR and/or 15 \mums and/or radio counterparts. For references see text.}
\tablehead{
\colhead{Seq Nr} & \colhead{RA} & \colhead{Dec} & 
\colhead{$J$}&	\colhead{$H$} & \colhead{$K$} & \colhead{{\it ISO} 15 \mums} &
\colhead{4.85 GHz} & \colhead{1.4 GHz} \\
\colhead{} & \colhead{} & \colhead{} & \colhead{} & \colhead{} & \colhead{} & \colhead{(mJy)} &
\colhead{(mJy)} & \colhead{(mJy)} }
\startdata
4 & 15:59:36.13&54:42:03.8& 16.61 $\pm$ 0.20 & 16.78 $\pm$ --   & 14.92 $\pm$ 0.16 & --              & --      & 4.20 $\pm$ 0.15   \\
12& 16:02:50.95&54:50:57.9&   --             &   --             &   --             & 1.99 $\pm$ 0.35 & --      & --                \\ 
17& 16:05:23.10&54:56:13.3&   --             &   --             &   --             & 2.02 $\pm$ 0.39 & --      & --                \\
18& 16:05:46.11&54:39:09.8& 16.90 $\pm$ 0.18 & 16.18 $\pm$ 0.22 & 15.57 $\pm$ 0.18 & --              &     171 & 80                \\
19& 16:06:23.56&54:05:55.8& 16.30 $\pm$ 0.14 & 16.14 $\pm$ --   & 15.37 $\pm$ 0.22 & 3.91 $\pm$ 0.63 & --      & 166.23 $\pm$ 0.146\\
20& 16:06:30.60&54:20:07.5&   --             &   --             &   --             & 3.14 $\pm$ 0.53 & --      & --                \\
21& 16:06:37.88&53:50:08.4&   --             &   --             &   --             & 1.46 $\pm$ 0.28 & --      & --                \\
22& 16:06:55.34&53:40:16.8& 16.37 $\pm$ 0.10 & 15.33 $\pm$ 0.10 & 14.32 $\pm$ 0.07 & --              & --      & --                \\
23& 16:07:05.16&53:35:58.5& 16.60 $\pm$ 0.12 & 16.29 $\pm$ 0.22 & 15.72 $\pm$ 0.22 & --              & --      & --                \\
27& 16:09:13.18&53:54:29.5&   --             &   --             &   --             & --              & --      & 43.82 $\pm$ 0.148 \\
31& 16:10:07.11&53:58:14.0&   --             &   --             &   --             & 1.59 $\pm$ 0.30 & --      & --
       \\
34& 16:12:38.27&53:22:55.0& 16.70 $\pm$ 0.15 & 16.15 $\pm$ 0.15 & 15.54 $\pm$ 0.23 & --              & --      & --                \\
\label{tabancdata}\enddata
\end{deluxetable}

\end{document}